\documentclass[10pt,letterpaper]{article}
\usepackage{opex3}
\usepackage{amsmath,amsthm,amsfonts,amssymb,verbatim,comment}
\usepackage[mathscr]{euscript}
\usepackage{graphicx}
\usepackage{appendix}
\usepackage[normalem]{ulem}
\usepackage{color,cancel}

\usepackage{ae} 

\def\sinc{{\rm sinc}}
\def\gb{\overline{\gamma}}

\begin{document}

\title{Theory of quantum frequency translation of light in optical fiber: application to interference of two photons of
different color}

\author{H. J. McGuinness,$^{1}$ M. G. Raymer,$^{1*}$ and C. J. McKinstrie$^{2}$}

\address{$^1$Department of Physics, University of Oregon, Eugene, OR 97403, United States \\
$^2$Bell Laboratories, Alcatel--Lucent, Holmdel, NJ 07733, United States}

\email{raymer@uoregon.edu} 

\begin{abstract}
We study quantum frequency translation and two-color photon interference enabled by the Bragg scattering four-wave 
mixing process in optical fiber. Using realistic model parameters, we computationally and analytically determine the 
Green function and Schmidt modes for cases with various pump-pulse lengths. These cases can be categorized as either 
``non-discriminatory" or ``discriminatory" in regards to their propensity to exhibit high-efficiency translation or 
high-visibility two-photon interference for many different shapes of input wave packets or for only a few input wave 
packets, respectively. Also, for a particular case, the Schmidt mode set was found to be nearly equal to a 
Hermite-Gaussian function set. The methods and results also apply with little modification to frequency conversion by 
sum-frequency conversion in optical crystals.
\end{abstract}


\bibliographystyle{osajnl}




\section{Introduction}

\begin{figure}[h] \vspace{-0.3in}
\centering\includegraphics[width=0.4\textwidth]{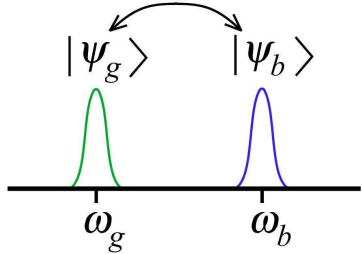} \vspace{-0.05in}
\caption{Quantum frequency translation of wavepacket states between green and blue spectral regions. The states being
translated could be coherent, squeezed, or number states.}
\label{fig:mikefig}
\end{figure}

Quantum frequency translation (QFT) is the process whereby two spectral modes (or narrow bands of modes) are swapped
in the sense that the properties of their quantum states (other than their central frequencies) are interchanged, as
in Fig. \ref{fig:mikefig} \cite{Vandevender,Tanzilli,McKinstrie2}. Such a process, of course, requires an external
source or sink of energy, which is provided by a strong external optical field. Ideally, an arbitrary state of the
narrow band of green modes will be swapped with the state of the narrow band of blue modes. These could be coherent,
squeezed, or number states, and may also be entangled with other, separate degrees of freedom, such as other modes
(e.g., violet and yellow) or states of atoms, quantum dots, etc. When the states of the green and blue modes are
swapped, the prior entanglement should be preserved. This paper focuses on the QFT of single-photon states.

QFT is defined to be ``noiseless," that is, free of excess background noise. For example, if both modes to be swapped
are initially vacuum, they
should remain vacuum$-$there are no spontaneous processes that can populate modes in this case. The noiseless
property distinguishes QFT from other frequency conversion processes such as optical parametric amplification,
which has been widely studied for use in classical telecom systems \cite{Gnauck,Mechin,McGuinness2}. In optical 
parametric amplification, power is
spontaneously transferred from the pump(s) to the weak signal fields, and without any blue input power, the signal
fields can build up spontaneously.

The QFT capability will play important roles in quantum information science (QIS), where states of a quantum memory
may be transmitted to a spatially remote quantum memory by emitting and absorbing photons. If the two memories
operate using distinct frequencies of their absorption/emission lines, it is necessary to ``translate" the
communication photons between these frequencies. Such a translation capability also allows sending the photons
through low-loss optical fiber at wavelengths near 1500 nm, even in cases where the memories operate at other
wavelengths.

Two methods for all-optical QFT have been demonstrated: three-wave mixing in a second-order nonlinear optical
medium (crystal) \cite{Tanzilli,Huang} and more recently four-wave mixing (FWM) in a third-order nonlinear optical
medium (silica fiber) \cite{McGuinness}. In the case of three-wave mixing and referring to Fig. \ref{fig:mikefig}, the 
strong
pump field must have frequency $\omega_p=\omega_b-\omega_g$. The configuration for FWM with two pumps, denoted $p$
and $q$, is that of dynamic Bragg scattering, in which for example, one photon is removed from pump $p$ and one is
added to pump $q$. At the same time, one photon is removed from the green signal and one is added to the blue signal
so that $\omega_p + \omega_g=\omega_q+\omega_b$. This contrasts with optical parametric amplification, wherein photons 
are removed from the
pump(s) and photons are added to both signals.

Experiments have included QFT of one-half of a number-correlated state \cite{Huang} and several examples of QFT of
single-photon states by three-wave mixing in crystals \cite{Vandevender,Tanzilli, Rakher} and by FWM in fiber 
\cite{McGuinness}.
The main advantage of the FWM method over the three-wave mixing method is that it allows QFT between spectral bands 
separated by
large or small shifts, whereas in practice three-wave mixing allows QFT only between far-separated bands 
\cite{McGuinness}. This
paper treats the theory of FWM, but the same methods can easily be applied to three-wave mixing.

A potentially revolutionary application of QFT is that it is predicted to enable the interference of two photons (or
other states of the modes) with different colors \cite{Raymer2}. This phenomenon generalizes the famous
Hong-Ou-Mandel (HOM) interference effect, wherein two photons of like color and identical wave-packet structure
impinge simultaneously on a standard beam splitter. Because of the boson nature of photons, the beam-splitter output
modes both contain either zero or two photons, but never one photon in each. This remarkable effect
(which is analogous to the exchange interaction for electrons) enables a class of quantum information processing
known as linear-optical quantum computing \cite{Knill}, including many variants \cite{Kok}.

The promise of QFT is that it offers the possibility to perform linear-optical quantum computing ``over the rainbow." 
That is, one could use the
light's color as a degree of freedom identifying optical qubits and perform linear-optical quantum computing using the 
generalization of the HOM
effect involving distinct frequencies. This would entail creating entangled states involving many distinct
frequencies (perhaps by methods such as in \cite{Menicucci, O'Brien}), and manipulating such states using two-mode
interference via QFT. This would require QFT between modes nearby in frequency, and so FWM would be greatly preferred
over three-wave mixing. Other tools proposed for linear-optical quantum computing ``over the rainbow" are temporal-mode 
filters (``pulse gates" \cite{Eckstein} which could be used as channel add/drop filters in an optical communications 
system \cite{Salehi, Marhic}) and single-photon pulse shapers \cite{Brecht,Kielpinski}, based on QFT by three-wave 
mixing.

In this paper we extend our earlier treatments \cite{McKinstrie2, Raymer2} to enable the explicit modeling of QFT by
FWM in optical fiber or by three-wave mixing in a crystal. The treatment includes analytical and numerical components, 
and allows us to model realistically the effects of finite pulse durations, linear dispersion, finite phase-matching 
bandwidth, and complicating effects such as self-phase modulation (SPM) and cross-phase modulation (CPM) \cite{HMdiss}. 
Previous theories of QFT have either assumed single-mode fields \cite{McKinstrie2}, disallowing rigorous treatment of 
pulse propagation, or assumed that the signal pulse shapes are not altered significantly by the translation process 
\cite{Eckstein, Brecht, Kielpinski}, obscuring the accurate design of high-fidelity devices.

The main results of this study are the following: We find that for realistic fiber and laser pulse parameters,
essentially 100 percent efficient QFT can, in principle, be achieved. We show that for given temporal shapes of the
pump pulses, usually there are unique shapes that the two signal wave packets should have in order that they be
swapped optimally by QFT. These shapes are found as the optimal modes of a singular-value (Schmidt) decomposition of
the Green function for the QFT process, and the shapes change significantly with changing translation efficiency.
This result provides a method for designing QFT and two-photon interference with high efficiency and high fidelity.
We also show, for a given pump temporal shape, how to design signal pulse shapes to create perfect HOM interference
via the FWM process, as first predicted in \cite{Raymer2}.

\section{Theory}

\begin{figure}[h]
\centering\includegraphics[width=0.4\textwidth]{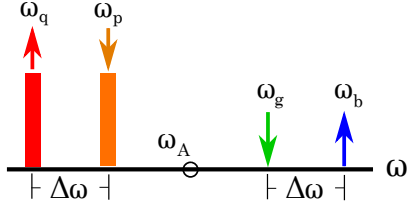}
\caption{BS process where the pumps have frequencies $\omega_p$ and $\omega_q$ and the signals have frequencies
$\omega_g$ and $\omega_b$, where $\omega_p + \omega_g = \omega_q + \omega_b$. The arrows symbolize either
annihilation (down) or creation (up) of a photon in that mode. The frequency $\omega_A$ is the average value of all
the frequencies and $\Delta \omega$ is the frequency separation between the pump fields and also between the signal
fields.}
\label{fig:BStoon}
\end{figure}

The physical mechanism used to carry out frequency translation  in fiber is the Bragg scattering (BS) four-wave
mixing process \cite{Gnauck,McGuinness,Inoue}, pictured in Fig. \ref{fig:BStoon}. For convenience, we refer to the
two weak signal fields as ``green" and ``blue," although in principle their actual frequencies could correspond to
visible or near-IR radiation. Two strong pump fields drive the process in which one photon from one of the pumps, pump
$p$
in Fig. \ref{fig:BStoon}, and one photon in the green field are annihilated while one photon from the other pump
(denoted $q$) and one photon in the blue field are created. The process conserves photon number in accordance with
the Manley-Rowe relations \cite{Boyd} and can be summarized as $\gamma_{p} + \gamma_{g} \rightarrow \gamma_{q} +
\gamma_{b}$, where $\gamma$ represents a photon.

Before describing the realistic continuous-frequency case, we discuss a much simpler model in which each field is
represented by a single-frequency mode. Such a model might apply to fields in a cavity, but strictly speaking not to
the pulse propagation scenario in which we are interested. The simple model has the advantage of permitting analytical
solutions, thereby providing some insight into the QFT process \cite{McKinstrie2}. For cases where each field can be
described by a single-frequency mode, the Hamiltonian governing the process is given by
\begin{equation}\label{heis}
\mathcal{H} = \Delta\beta (a^\dagger_g a_g - a^\dagger_b a_b)/2 + \kappa a^\dagger_g a_b + \kappa^* a^\dagger_b
a_g,
\end{equation}
where the creation/annihilation operators satisfy the usual commutation relations, and $\Delta \beta$ and $\kappa$
are parameters quantifying the effects of phase-mismatch and nonlinearity \cite{McKinstrie2}. The evolution is given
by the spatial equations-of-motion (with a minus sign compared to the usual time-domain Heisenberg equation)
\begin{equation}\label{eq:Heiseq}
\frac{d}{dz} a_{b(g)}(z) = i[a_{b(g)}(z),\mathcal{H}].
\end{equation}

The solution of \eqref{eq:Heiseq} for $a_{g(b)}(z)$ as a function of position along the fiber $z$ for a fiber of
length $L$ is given by
\begin{align}
\label{eq:CWBS1}
a_g(z) & = \mu(z)a_g(0) + \nu(z)a_b(0), \\
\label{eq:CWBS2}
a_b(z) & = \mu^*(z)a_b(0) - \nu^*(z)a_g(0),
\end{align}
where the ``transfer" functions are given by
\begin{align}
\label{eq:trans1}
\mu(z) & = \textrm{cos}(k z) + i\Delta \beta \textrm{sin}(k z)/2k, \\
\label{eq:trans2}
\nu(z) & = i \kappa \; \textrm{sin}(k z)/k,
\end{align}
where $k = (\Delta\beta^2/4 + |\kappa|^2)^{1/2}$. The transfer functions have the relation $|\mu|^2 +|\nu|^2 =
1$,
which ensures the process is unitary and conserves photon number. Mathematically, the BS translation process having
frequency input and output ``ports" is identical to the normal beam-splitter operation on degenerate monochromatic
fields \cite{McKinstrie3}. For perfect translation from green to blue, and vice
versa, it must be that $|\nu(L)|=1$ for a medium of length $L$. In this case the quantum states of the modes are 
completely swapped. For example, a single green photon is replaced by a single blue photon, that is,  $|1,0\rangle 
\rightarrow |0,1\rangle$, where $|i,j\rangle$ denotes the state in which $i$ photons are in the green region and $j$ 
photons are in the blue region.

Since the BS process is analogous to beamsplitter action, all of the effects that a beamsplitter can induce on a
field will also occur in the BS process. In particular, the effect analogous to two-photon HOM
interference will occur, where two independent photons of \emph{differing frequencies} will interfere with one
another and give a perfect HOM dip under the appropriate circumstances \cite{Raymer2}. Just as in the monochromatic
HOM interference effect, the ``frequency beamsplitter" must have equal transmission and reflection coefficients, that
is $|\mu|^2 = |\nu|^2 = 1/2$, so that it is equally likely for the blue (green) photon to stay in the blue (green)
mode or be translated to the green (blue) mode. Using the above equations it can be shown that in this case, for an
input state consisting of one green photon and one blue photon, denoted by $
a_g^{\dagger}(0)a_b^{\dagger}(0)|\rm{vac}\rangle = |1,1\rangle_{\rm{in}}$, the output state is given by
\begin{equation}
\label{eq:CWhomi}
|1,1\rangle_{\rm{in}}  = \left( [|\mu|^2 - |\nu|^2] |1,1\rangle_{\rm{out}} + \sqrt{2}\mu \nu|2,0\rangle_{\rm{out}} -
\sqrt{2}\mu^* \nu^*|0,2\rangle_{\rm{out}}  \right).
\end{equation}
For equal values of $|\mu|^2$ and $|\nu|^2$ the $|1,1\rangle$ component of the state cancels to zero.

The above analysis is valid only if all the fields involved are single-mode. For fields that contain
multiple-frequency components, as any pulsed field does, the theory becomes much more complex. For arbitrary input,
and even ``simple" multi-mode input such as Gaussian pulses, the output cannot in general be determined analytically.
However, it is possible to assign analytically a Green function $G$ in either frequency $\omega$ or time $t$ that
relates the input wavepacket to the output wavepacket for arbitrary input, analogous to \eqref{eq:CWBS1} and
\eqref{eq:CWBS2}, for given pump spectra, fiber dispersion, and fiber length \cite{Raymer2}. Denoting the creation
operator at position $z$ by $a^{\dagger}(z,\omega)$, the input $a^{\dagger}(0,\omega)$, and output
$a^{\dagger}(L,\omega)$ are related by the \emph{forward} transformation
\begin{equation}\label{eq:Green'sfor}
a^{\dagger}(L,\omega) = \int d\omega^{'} G^*(\omega^{'},\omega)a^{\dagger}(0,\omega^{'}).
\end{equation}
For our purposes, the \emph{backward} transformation is more useful:
\begin{equation}\label{eq:Green's}
a^{\dagger}(0,\omega) = \int d\omega^{'} G(\omega,\omega^{'})a^{\dagger}(L,\omega^{'}),
\end{equation}
which gives the input operator in terms of an integral over the output operator.

A single-photon input wavepacket given by $\int d\omega A(0,\omega) a^{\dagger}(0,\omega)|vac\rangle$, where 
$A(0,\omega)$ is the spectrum of the input wavepacket, is translated in the manner given by \eqref{eq:Green's}. The 
state can thus be expressed in terms of the output operator as
\begin{equation}\label{eq:AinGreen}
|\psi\rangle = \int d\omega A(0,\omega) \int d\omega^{'}
G(\omega,\omega^{'})a^{\dagger}(L,\omega^{'})|vac\rangle.
\end{equation}

The Green function is unitary and therefore satisfies the condition
\begin{equation}\label{eq:unitary}
\int d \omega^{'} G(\omega,\omega^{'})G^{*}(\omega^{''},\omega^{'}) = \delta(\omega-\omega^{''}).
\end{equation}

To facilitate an intuitive understanding of the transfer function, we break up frequency space into two pieces; the
range that contains the green and the range that contains the blue. Each operator or variable will be denoted by a
``$g$" or ``$b$" subscript. For example, the creation operator for an green photon at a particular green frequency
will be written as $a^{\dagger}_{g}(\omega_{g})$. In this case the transfer function can be written in block form as
\cite{Raymer2}
\begin{equation}
\left[ \begin{array}{cc} G_{gg}(\omega_{g},\omega_{g}^{'}) & G_{gb}(\omega_{g},\omega_{b}^{'})\\
G_{bg}(\omega_{b},\omega_{g}^{'}) & G_{bb}(\omega_{b},\omega_{b}^{'})
\end{array} \right]
\end{equation}
where $G_{xy}$ refers to the evolution of an output creation operator in range $y$ which came from range $x$.


Although the $G$ function completely describes the evolution of arbitrary inputs, it is difficult to discover which
input pulses give rise to desirable effects such as high-efficiency translation or good two-color HOM interference
just by examining the $G$ function. However, by performing a Schmidt (singular-value) decomposition (SVD) \cite{Ekert} 
on the $G$ function or on the sub-matrices $G_{xy}$, the input wavefunctions necessary to achieve these effects can be 
found \cite{Raymer2}. For example, consider the translation of an input green field to the blue field, which is 
described by the Green sub-matrix by $G_{gb}$, which is not Hermitian and therefore does not admit a unitary 
decomposition. We write $G_{gb}$ in its SVD form as
\begin{equation}\label{eq:GgbSVD}
G_{gb}(\omega, \omega^{'}) = -\sum_n \rho_n V_n(\omega) w_n^{*}(\omega^{'}),
\end{equation}
where the $\rho_n$ are the real, positive singular values (the generalization of eigenvalues), and $V_n$ and $w_n$
are the $n$th column vectors of the unitary ``matrices" $V$ and $w$ describing the decomposition,
respectively. The most useful way to interpret this decomposition is that a given green wavepacket $V_n$ will be
translated to the blue wavepacket $w_n$ with probability $\rho_n^2$. Hence, only input green wavepackets $V_n$
that have associated singular values $\rho_n=1$ will translate over to the blue mode with 100 percent probability.
The $G_{gg}$ ``matrix" describes an input green wavepacket scattering within the green spectral region at the output
(no translation from green to blue, although the shape of the wavepacket may change). If the SVD of $G_{gg}$ is given
by
\cite{Raymer2}
\begin{equation}\label{eq:GggSVD}
G_{gg}(\omega,\omega^{'}) = \sum_n \tau_n V_n(\omega) \upsilon_n^{*}(\omega^{'}),
\end{equation}
then $\tau_n^2$ is the probability that an input green wavepacket $V_n$ will scatter to an output green wavepacket
$\upsilon_n$ (the $V$ matrix here being the same as in the $G_{gb}$ decomposition). The singular values satisfy the
relation $\tau_n^2 + \rho_n^2 =1$, as they must to conserve photon number.


Similarly, the $G_{bg}$ and $G_{bb}$ have their own SVDs, given by
\begin{align}
G_{bg}(\omega, \omega^{'}) & = \sum_n \rho_n W_n(\omega) \upsilon_n^{*}(\omega^{'}), \label{eq:GbgSVD} \\
G_{bb}(\omega, \omega^{'}) & = \sum_n \tau_n W_n(\omega) w_n^{*}(\omega^{'}), \label{eq:GbbSVD}
\end{align}
where the $w$ and $\upsilon$ matrices, and the singular values $\tau$ and $\rho$, are the same as in
\eqref{eq:GgbSVD} and \eqref{eq:GggSVD}. This highlights the interconnectedness between the various sub-Green
functions in that there are only four unique Schmidt modes for each index $n$ that completely describe the process. 
Before treating exclusively single-photon inputs, we emphasize that the Green functions determined by our present 
approach can be used to treat arbitrary multi-photon inputs as well.

The SVD technique can be used to discover what input wavepackets are suitable for optimal two-photon HOM
interference \cite{Raymer2}. The condition is somewhat intuitive. Since in the single-mode, two-color case it is
necessary for both green and blue inputs to have a 50 percent probability of translating to the other mode for
perfect HOM interference, it is reasonable to guess this condition is also true in the multi-mode two-color case.
Hence for the cases in which $\tau_n^2 = \rho_n^2 =1/2$, the corresponding input vectors $V_n$ and $W_n$ will show
perfect two-color HOM interference. Alternatively, this condition can be written as
\begin{equation}\label{eq:HOMIsvd}
\sigma_n = 2\tau_n \rho_n=1,
\end{equation}
We refer to the quantity $\sigma_n = 2\tau_n \rho_n$ as the HOM singular value, because it appears as such for a
Schmidt decomposition kernel designed specifically to optimize the HOM interference. (See Eq. (37) of \cite{Raymer2}.)
The closer it is to unity, the better the HOM interference.


There is an alternative, complimentary, method to calculate the degree of HOM interference that a BS process will
exhibit, for which the Green function does not have to be calculated. First, consider the case in which there are two
photons, either of which could be in the green or blue region, or in some superposition of both regions. In general,
the state of this bi-photon can be written as
\begin{equation}\label{}
| \psi \rangle = C_{11}|1,1\rangle + C_{20}|2,0\rangle + C_{02}|0,2\rangle.
\end{equation}
The entire state $| \psi\rangle$ must be normalized and the component states are orthonormal. Now, for input
consisting of one green photon and one blue photon having input spectra of $A_{g}(0,\omega_g)$ and
$A_{b}(0,\omega_b)$ respectively, the state is
\begin{equation}\label{}
\int d\omega_g d\omega_b A_{g}(0,\omega_g) A_{b}(0,\omega_b)a_g^{\dagger}(0,\omega_g)
a_b^{\dagger}(0,\omega_b)|vac\rangle.
\end{equation}
If the photons undergo the BS process, the input photon operators evolve according to \eqref{eq:Green's}. By switching 
the limits of integration and noting that $\left[\int d\omega A_{X}(0,\omega)G^*_{XY}(\omega,\omega_{Y}) \right]
\rightarrow  A_{XY}(L,\omega_{Y})$, where both $X$ and $Y$ could be either $g$ or $b$ and $A_{XY}(L,\omega_Y)$ is a
component of the spectrum at the output of the fiber, the state can be written compactly as
\begin{equation}\label{eq:HOMIout2}
\begin{split}
    |\psi \rangle =
    & \big[ \int d\omega_{g} a_{g}^{\dagger}(L,\omega_{g}) A_{gg}(L,\omega_{g}) + \int
d\omega_{S}a_{b}^{\dagger}(L,\omega_{b}) A_{bg}(L,\omega_{b}) \big] \\
    & \times \big[ \int d\omega_{g}^{'} a_{g}^{\dagger}(L,\omega_{g}^{'}) A_{gb}(L,\omega_{g}^{'}) + \int
d\omega_{b}^{'} a_{b}^{\dagger}(L,\omega_{b}^{'}) A_{bb}(L,\omega_{b}^{'}) \big]|vac \rangle.
\end{split}
\end{equation}

The probability, given a particular output state, for the various sub-space states to occur is given by the modulus
squared of the coefficients of these states, $|C_{ij}|^2$. For the $|1,1\rangle$ state the coefficient is found by
acting the aforementioned operators, with their accompanying functions, on the vacuum, yielding
\begin{equation}\label{}
| \psi_{11} \rangle = \int \int  d\omega_{g} d\omega_{b} \left[ A_{gg}(L,\omega_{g}) A_{bb}(L,\omega_{b}) +
A_{gb}(L,\omega_{g}) A_{bg}(L,\omega_{b}) \right] |1_{\omega_{g}}, 1_{\omega_{b}}\rangle
\end{equation}
The first term, $A_{gg}(L,\omega_{g}) A_{bb}(L,\omega_{b})$, corresponds to the case where a photon initially in the
green mode (i.e. wavepacket) scatters within the green mode at the output, and a photon initially in the blue mode
scatters within the blue mode at the output. The second term, $A_{gb}(L,\omega_{g}) A_{bg}(L,\omega_{b})$,
corresponds to the case where a photon initially in the green mode scatters to the blue mode at the output, and a
photon initially in the blue mode scatters to the green mode at the output. Since these terms are being added before
their modulus squared is taken, there is the chance that they could cancel if the terms are proportional and out of
phase with one another. This is exactly what happens in the case in which ideal HOM interference occurs, leaving zero
probability
amplitude for the field to emerge in the $|1,1\rangle$ state. The expression for the probability for this state to
occur $P_{11}$ is
\begin{equation}\label{eq:P11}
\begin{split}
    P_{11} = \int \int  d\omega_{g} d\omega_{b} & \big[ |A_{gg}(L,\omega_{g})|^2 |A_{bb}(L,\omega_{b})|^2 +
|A_{gb}(L,\omega_{g})|^2 |A_{bg}(L,\omega_{b})|^2 +\\
     & 2\mathrm{Re}[ A_{gg}(L,\omega_{g})A_{bb}(L,\omega_{b})A_{gb}^*(L,\omega_{g}) A_{bg}^*(L,\omega_{b})|] \big].
\end{split}
\end{equation}

This gives an expression to calculate the coincidence probability for any given input pulses without knowing the
Green function for the process. In the alternative case in which we launch particular Schmidt modes as input signals,
we showed previously that the coincidence count rate equals $P_{11} = (\tau_n^2 - \rho_n^2)^2$ \cite{Raymer2}.
Therefore, in this case we have a simple relation between $P_{11}$ and the HOM singular values; $P_{11} +
\sigma_n^2=1$. Other important quantities are $P_{20}$ and $P_{02}$, the probabilities that two photons appear in one or 
the other output ports together. The expressions for these, and their derivation can be found in Appendix A.

\section{Numerical implementation}

Unfortunately, it is not possible to discover the Green function and its Schmidt modes by examining the Hamiltonian 
directly. One must solve for the full propagation first, then examine the input-output relations. This emphasizes that 
the Schmidt modes are global properties, depending on the entire evolution. For example, the input Schmidt modes depend 
on the length of the medium.

The appropriate Hamiltonian generates the well-known nonlinear Schrodinger equation, which needs to be solved. In most 
realistic circumstances, taking into account actual field shapes and real fiber dispersion properties, it is not 
possible to solve the nonlinear Schr$\ddot{\textrm{o}}$dinger equation for the fields analytically. So we turn to
numerical computation to determine the evolution of a completely specified input state to a completely determined
output state. The key difficulty with such an approach is that, in general, nonlinear operator equations cannot be
simulated by any simple numerical scheme. When the evolution is both in space and time as in our case, the problem
becomes one suited for techniques such as the positive-P representation \cite{Drummond, Gardiner}. Fortunately, in
our case, although the evolution is nonlinear in the pump fields, it is linear in the weak signal fields. This allows
a simplification if we treat the pump fields as classically evolving parameters in the quantum Heisenberg equations
of motion  for the weak signals. Then the field operators for the weak signals can be solved in terms of Green
functions that obey the classical evolution equations.  Therefore, we solve numerically for the classical evolution
of the Green functions and use those solutions to calculate any quantum probability or correlation function
concerning the signal fields. An analogous procedure was used in \cite{Wasilewski1,Wasilewski2}.

The usual starting point for solving for the nonlinear evolution of a pulse is to write the total field as $E(z,t) =
A(z,t)e^{i(\beta_0 z-\omega_0 t)}$, where $A(z,t)$ is a slowly varying envelope function, $\omega_0$ is the high
frequency of the carrier wave, and $\beta_0$ is the propagation constant at $\omega_0$. In that form the evolution is
governed by the nonlinear Schr\"odinger equation
\begin{equation}
\partial_z A(z,t) = i\beta(i\partial_t)A(z,t) +
i\gamma |A(z,t)|^2 A(z,t).
\end{equation}
Although this is effective for single pulses, for electric fields made up of multiple,
distinct pulses with significantly different frequency components, solving this form of the equation is unnecessarily
computationally intensive and impractical. Instead, we solve four coupled equations, derived from the nonlinear
Schr\"odinger equation, in which terms relating to dispersion, self-phase modulation, cross-phase, and
the BS process were kept. Numerically, the frequency components of each field are associated with a frequency mesh
pertaining only to that field. This drastically reduces the number of calculations needed to compute the solution in
that it is unnecessary to resolve the large empty bandwidth in between fields. In this scheme, the total electric
field is the sum of the four individual fields, each having a slowly varying amplitude function $A_j$ centered around
a carrier wave of frequency $\omega_j$;
\begin{equation}\label{eq:Efield4A}
E(z,t) = \sum_j A_j(z,t)e^{i(\beta_j z-\omega_j t)} + \rm{c.c.},
\end{equation}
where $j$ denotes $p$, $q$, $g$ or $b$. In quantum theory, $A_j$ is proportional to an annihilation operator. 
Substituting this ansatz into the nonlinear Schr$\ddot{\textrm{o}}$dinger equation leads to, with suppressed independent 
variables $z$ and $t$, the four coupled BS equations (where $\partial_z$ represents the partial derivative with respect 
to $z$)
\begin{equation}\label{eq:4mode}
\begin{split}
\partial_z A_p &= i\beta_{p}(i\partial_T; \omega_p)A_p  + \left(i\gamma |A_p|^2 + 2i\gamma \sum_{k \neq p} |A_k|^2
\right) A_p + 2i\gamma A_g^* A_q A_b \mathrm{exp}[-i \Delta \beta z]\\
\partial_z A_q &= i\beta_{p}(i\partial_T; \omega_q)A_q  + \left(i\gamma |A_q|^2 + 2i\gamma \sum_{k \neq q} |A_k|^2
\right) A_q + 2i\gamma A_b^* A_p A_g \mathrm{exp}[i \Delta \beta z]\\
\partial_z A_g &= i\beta_{p}(i\partial_T; \omega_g)A_g  + \left(i\gamma |A_g|^2 + 2i\gamma \sum_{k \neq g} |A_k|^2
\right) A_g + 2i\gamma A_p^* A_q A_b\mathrm{exp}[-i \Delta \beta z]\\
\partial_z A_b &= i\beta_{p}(i\partial_T; \omega_b)A_b  + \left(i\gamma |A_b|^2 + 2i\gamma \sum_{k \neq b} |A_k|^2
\right) A_b + 2i\gamma A_q^* A_p A_g \mathrm{exp}[i \Delta \beta z],
\end{split}
\end{equation}
where $\Delta \beta$ is the dispersive mismatch given by $\Delta \beta = \beta_p + \beta_g - (\beta_q +\beta_b) +
\gamma(P_q - P_p)$, $P_q$ and $P_p$ are the peak pump powers, and $\gamma$ is the nonlinear coefficient. The
first term on the right hand side describes the effect of linear dispersion, the next two terms describe SPM and CPM
respectively, and the last term describes the BS four-wave mixing process. The quantity
$\beta_{p}(i\partial_T; \Omega_0)$ is given by
\begin{align}
\label{eq:beta_eq3}
\beta_{p}(i\partial_T; \Omega_0) &= \sum^\infty_{n=1} \frac{\beta^{(n)}(\Omega_0) (i\partial_T)^n}{n!} -
\beta^{(1)}(\omega_p)i\partial_T.
\end{align}
where $\beta^{(n)}(\Omega_0)$ is the $nth$ derivative of $\beta$ evaluated at $\Omega_0$, and $T = t -
\beta^{(1)}(\omega_p)z$. In this scheme all the fields are propagating in the arbitrary choice of the frame of
reference of pump $p$. With one frame of reference there need only be one universal time mesh.


Although \eqref{eq:4mode} is for quantum field operators, we can realistically replace the pump fields $A_p$ and
$A_q$ by classical functions if these fields are in strong coherent states. We solve the first two equations for the
pumps ignoring the small effects of the weak signal fields $A_g$ and $A_b$, but retaining the SPM and CPM effects of 
the
strong pumps. We then use these pump solutions in the second pair of equations for the
weak signal fields. Because these equations are linear in the operators, we can treat them formally as classical
equations. Then the quantum effects can be fully accounted for by using the commutation relations when calculating
quantities such as $P_{11}$, as describe above.

To solve the coupled equations numerically we use the split-step Fourier method (SSFM) \cite{Agrawal,Press}. For each
electric field $E_j$ we approximately solve the operator equation $\partial_z A_j(T,z) = (\hat{D}_j + \hat{N}_j)
A_j(T,z)$, where $\hat{D}_j$ and $\hat{N}_j$ are the dispersive and nonlinear differential operators, acting in the
frequency and time domains respectively, which act on $E_j$ given in Eq. \eqref{eq:4mode}. The approximation is that
$A(T,z)$ is taken to $A(T,z + \Delta z)$ by
\begin{equation}\label{eq:SSSFev}
A(T,z + \Delta z) \approx \mathrm{exp}\left(\frac{\Delta z}{2}\hat{D}\right) \mathrm{exp}\left( \int_{z}^{z+\Delta z}
\hat{N} dz \right) \mathrm{exp}\left(\frac{\Delta z}{2} \hat{D}\right)A(T,z).
\end{equation}
where $\Delta z = L/N$ and $N$ is the number of steps. The nonlinear component integral is found by use of a
fourth-order fixed-step Runge-Kutta method.


The SSFM allows one to calculate a particular output amplitude for a particular arbitrary input amplitude. However,
discovering the Green function that governs the process requires solving the propagation equation for a family of
orthogonal input amplitudes. Let $A_{in}$ be an $n \times n$ matrix with each column equal to one time-domain vector
from an orthogonal set, $A_{out}$ be an $n \times n$ matrix with each column equal to the output of the $n^{th}$
column of $A_{in}$, and $G$ be the Green function. In this paper we use the Hermite-Gaussian (HG) functions as the
orthogonal set, where the $n^{th}$ function is given by
\begin{equation}
\psi_n(x)=\left(\frac{1}{2^n n! \sqrt{\pi}} \right)^{1/2} e^{-x^2/2} (-1)^n e^{x^2}
\frac{d^n}{dx^n}e^{-x^2},
\end{equation}
where the characteristic time of the basis set is set by the overall time window of the mesh. Due to the linear
nature of the Green function formalism, as expressed in \eqref{eq:Green's}, the \emph{forward} Green function expressed 
in matrix notation is
\begin{equation}\label{eq:simpG}
G^{\dagger} = A_{out} \cdot A_{in}^{-1}.
\end{equation}
This is strictly true only if all of the output vectors of $A_{out}$ can be described as some linear combination of
the orthogonal vectors of $A_{in}$. In practice, an orthogonal vector of order $j$ will evolve to have components of
both lower and higher order than $j$. In the cases where the $j$th mode evolves to include components of higher
order than $n$, $G$ won't satisfy the unitary conditions of \eqref{eq:unitary}, which is often the case for the
evolution of the very-high-order modes. In that case, only the subset of orthogonal input vectors that evolve into
output that can be described by all $n$ input vectors can be considered ``valid" output, as far as determining the
Green function is concerned. Let $A_{in,sub}$ be a $m \times m$ dimensional matrix  that is the sub-matrix of $A_{in}$ 
where all the input vectors evolve to output vectors (described by the $n \times m$ matrix
$A_{out,sub}$, where $m\leqq n$) that can be described as a linear combination of the vectors of $A_{in}$. Then the 
Green function that
always satisfies the unitary conditions is given by
\begin{equation}\label{eq:Gsub}
G_{sub}^{\dagger} = A_{out,sub} \cdot A_{in,sub}^{-1},
\end{equation}
and will be a matrix of dimensions $n \times m$. In practice, the simplest way in which to determine the correct
matrix subsets is by applying the unitary conditions (calculating $G \cdot G^\dagger$) and seeing for what subset of
vectors they hold. When that is determined, $G_{sub}^{\dagger}$ can then be calculated.

\begin{figure}[h]
\centering\includegraphics[width=0.5\textwidth]{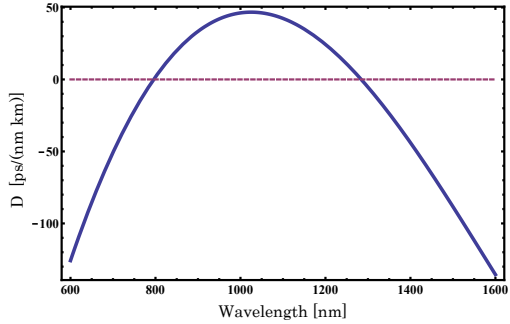}
\caption{The $D$ parameter for the photonic crystal fiber used in simulations from 600 nm to 1600 nm. The dashed line
is at $D =0$ for reference.}
\label{fig:fibdis}
\end{figure}

Finally, there is the question of fiber modeling. As these calculations are meant to help explain past experimental
results \cite{McGuinness} and guide future experiments using solid-core photonic crystal fiber \cite{Russell},
we desire a fiber model that fairly accurately describes photonic crystal fibers. We employ the often used step-index 
model in describing photonic crystal fibers \cite{Wong}, which approximates the photonic crystal fiber as a step-index 
fiber having a core index $n_{core}$ (usually the index of bulk fused silica) and an effective cladding index 
$n_{clad}$, where $n_{clad}(\omega) = f + (1 - f) n_{core} (\omega)$. Here $f$ is the air-filling fraction of the 
lattice region compromising the cladding, which along with the core radius $a$, parameterizes the fiber. By measuring 
the modulation instability wavelength signals produced by a fiber as a function of pump wavelength, best-fit values for 
$f$ and $a$ can be determined. For a particular fiber that has been used in past experiments and is a good candidate for 
future experiments, this procedure has been carried out, resulting in parameter values of $f=0.494$ and $a=0.72$ $\mu$m, 
with the resulting dispersion parameter $D$ shown in Fig. \ref{fig:fibdis}. This is the dispersion profile used for all 
numerical calculations in this paper.

\begin{figure}[h]
\centering\includegraphics[width=2.8in]{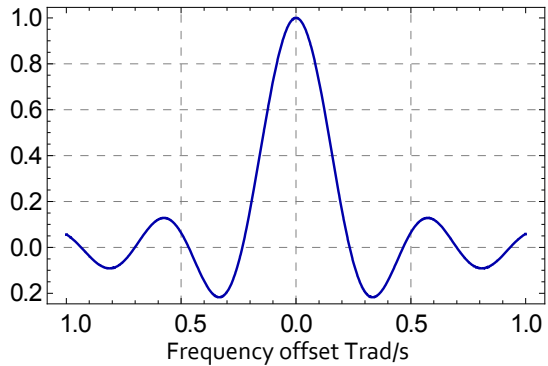}
\caption{Phase-matching function sinc($\Delta \beta L/2)$ for the fiber used in the simulation, where the pump 
wavelengths were fixed and the ``green" and ``blue" fields frequencies were varied in an energy-conserving manner. The 
horizontal axis is the ``green" frequency variation from its corresponding central wavelength value of 673 nm. For this 
plot the fiber length is $L=20$m, and the FWHM of the function is approximately 0.3 Trad/s ($10^{12}$ rad/s).}
\label{fig:fiberPM}
\end{figure}

Given a particular dispersion profile of a fiber and the central wavelengths at which translation will occur, the
fiber can effectively translate input signals only with spectral widths smaller than the characteristic bandwidth of
the BS process. As seen in equations \eqref{eq:trans2} in the case of low conversion efficiency, the state is
proportional to sinc$(\Delta \beta L/2)$, which defines the dispersive mismatch of the process. Figure
\ref{fig:fiberPM} shows this sinc function for the fiber, central wavelengths, and length (20 m) used in the
simulations. All fields were taken to be monochromatic, and the pumps where fixed at 808 nm and 845 nm. The ``blue"
signal field was varied from its central wavelength of approximately 649 nm (shown in frequency on the horizontal
axis) while the ``green" field was varied from its central wavelength of 673 nm to conserve the energy of the process
$(\omega_s = \omega_p +\omega_i - \omega_q)$. The full-width at half-maximum (FWHM) of the central sinc lobe is
approximately 0.3 Trad/s. This number gives a rough estimate of the effective bandwidth of the BS process. As we will
see, all else being equal, input signals that have spectral widths larger than this number do not translate as
efficiently as signals that have spectral widths lower than this number.

\section{Numerical results}

This section details the results of simulations for several different cases, broken up into two main categories;
cases that highlight properties relating to effective total translation from one signal field to another, and
cases that highlight properties relating to good two-color HOM interference. For the cases in which the Green
function of the process was calculated, the two pumps were identical transform-limited Gaussian-shaped fields. The
pump \emph{intensity} profiles had either a ``long" FWHM duration of 1000 ps, making them quasi-CW for most input
signal fields used, or a ``short" FWHM duration of 70 ps. At the input to the fiber, the centers of the fields were
placed at zero time. The central wavelengths for the pump fields $p$ and $q$ were 808 nm and 845 nm,
respectively, placing them in the anomalous-dispersion region. The signals ``green" $g$ and ``blue" $b$ were at
wavelengths 673 nm and 649 nm, respectively, placing them in the normal-dispersion region. The 845-nm and 649-nm fields 
had nearly equal group velocities ($\cong 2.0044 \times 10^8$ m/s), and the 808-nm and 673-nm fields had nearly equal 
group velocities ($\cong 2.0049 \times 10^8$ m/s). The nonlinear coefficient $\gamma$ was set to $100 \; 
\textrm{W}^{-1}\textrm{km}^{-1}$ for all calculations.

The Green functions were calculated in the HG mode basis, where the input and output field amplitudes were decomposed 
into a common set of HG functions. The Green function for a particular case was found by sequentially setting the green 
or blue input field to a single Hermite-Gaussian function, solving the propagation, and projecting the output pulse onto 
the HG basis functions. Thereby the green function is represented in the HG mode basis. The characteristic time scale 
for the HG functions, which we define as the FWHM of the zeroth order HG function, used to produce the figures was found 
by first arbitrarily choosing a characteristic time of 40 ps, fitting the resulting absolute value of the first Schmidt 
mode to a Gaussian function\footnote{In most cases the first Schmidt mode was quite Gaussian-like.}, and calculating the 
Green function again using this new characteristic time. This time-scale choice for defining an ``optimal HG basis" 
produced Green function matrices that were the least complex, some of which were nearly diagonal, and therefore the 
simplest to understand. The HG basis set was centered in time at the peak of the pump and signal inputs.

Parameter studies were also carried out in which the two pump pulses' intensity durations were set at 1000 and 70
ps, and the input signal field durations were varied. Either the translation efficiency or HOM interference properties 
were documented, depending on the main category of the case.

There are a few ways in which to gauge the accuracy of the numerical solutions. While not explicitly built into the 
calculations, for all calculations the sum of the green and blue pulse energies (calculated in the frequency
domain) were found to be conserved, in accordance with the Manley-Rowe relation \cite{Boyd} for the BS process. Also,
there are unitary conditions on the Green functions that govern the process, given by \eqref{eq:unitary}. Although 
checking these conditions is significantly more expensive computationally than verifying the Manley-Rowe relations, and 
therefore is not feasible for studies requiring large numbers of cases like the below-mentioned parameter studies, this 
check was carried out for several representative cases for such studies. The Green functions were then checked against 
\eqref{eq:unitary}, and in these cases were found to agree well. Finally, the appropriate number of steps were taken for 
each calculation so that the solutions converged and were not meaningfully dependent on the exact number of steps.

\subsection{Frequency translation, long pulse} \label{sec:LongPHighP}

\begin{figure}[h]
\centering\includegraphics[width=0.8\textwidth]{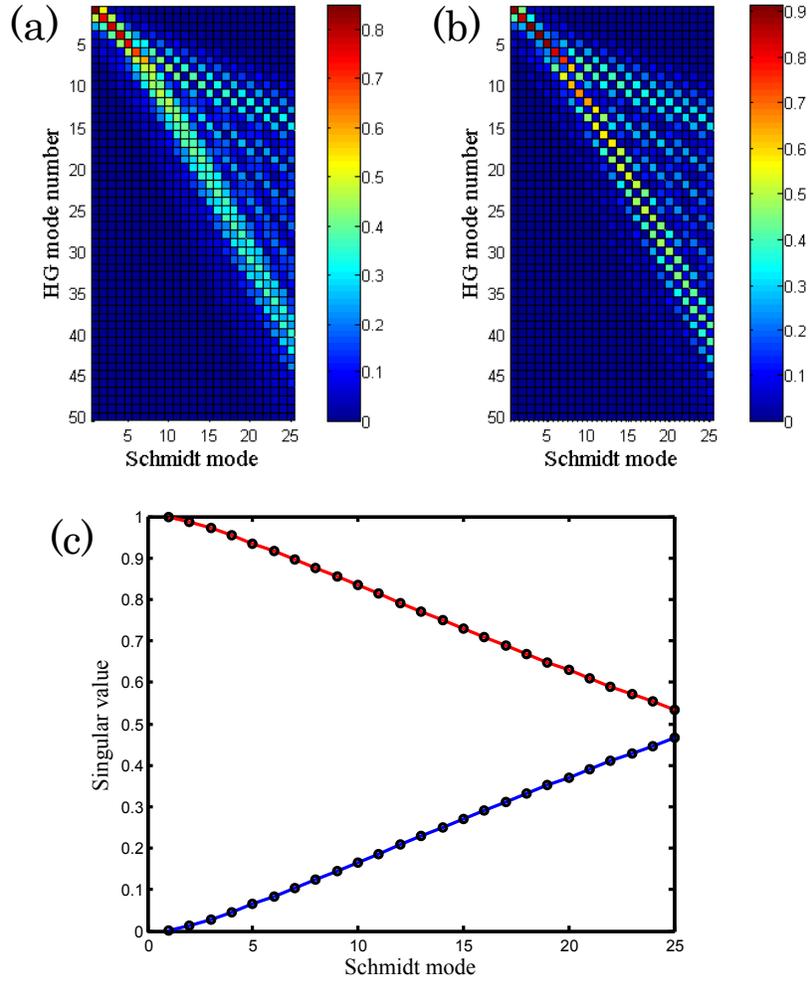}
\caption{Singular values, $V$ matrix (columns of which are coefficients in the HG basis for individual Schmidt modes), 
and $w$ matrix of $G_{gb}$ Green function relating to the case with long pumps
(1000 ps) having peak powers of 400 mW and a fiber length of 20 m. (a) Absolute value of the $V$ matrix (input
Schmidt modes). (b) Absolute value of the $w$ matrix (output Schmidt modes). (c) Squared singular values $\rho^2$
(red) and $\tau^2$ (blue) indexed by the corresponding Schmidt mode number.}
\label{fig:285.1-3}
\end{figure}
\begin{figure}[h]
\centering\includegraphics[width=0.5\textwidth]{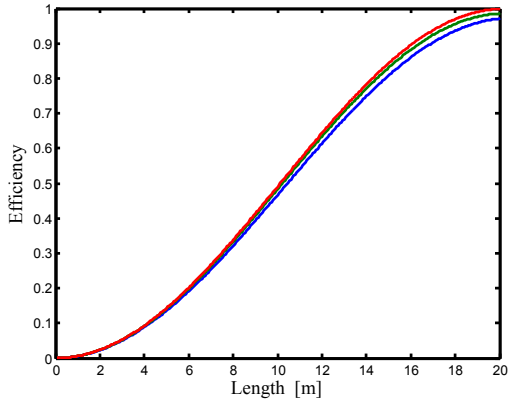}
\caption{Translation efficiency as a function of length along the fiber relating to the case with long pumps (1000
ps) having peak powers of 400 mW. The red, green, and blue lines relate to the first, second, and third green Schmidt
modes, respectively.}
\label{fig:285.4}
\end{figure}
\begin{figure}[h]
\centering\includegraphics[width=0.9\textwidth]{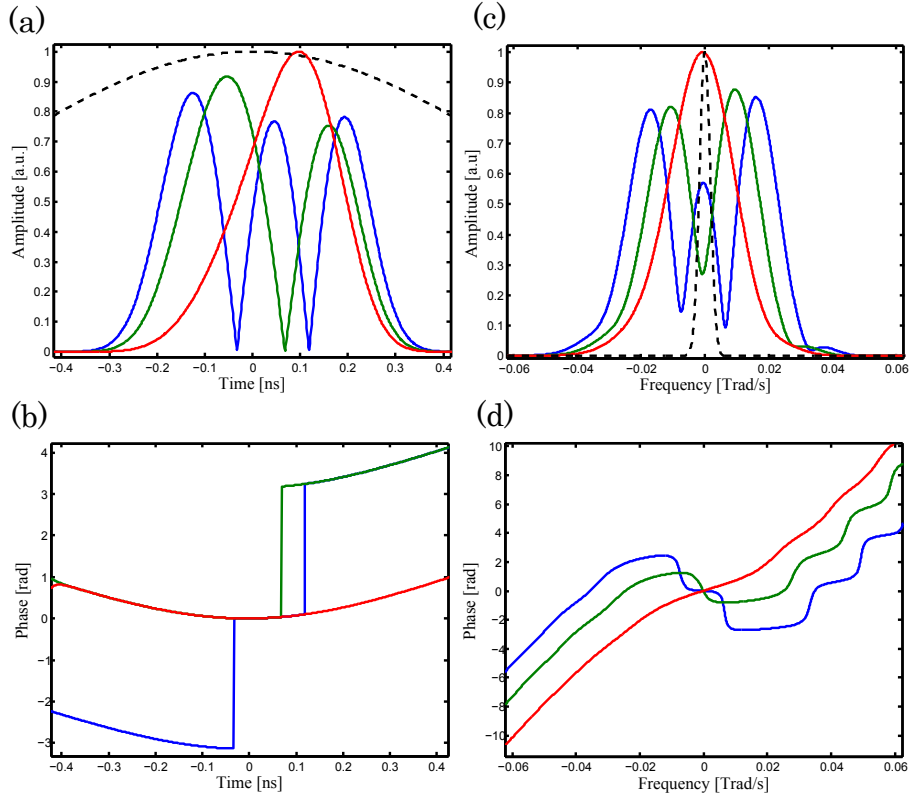}
\caption{Amplitude and phase in both frequency and time of the green 673-nm input Schmidt modes relating to the case 
with long pumps (1000 ps) having peak
powers of 400 mW. The red, green, and blue lines relate to the first, second, and third Schmidt modes, respectively.
Note: Trad/s indicates $10^{12}$ rad/s.}
\label{fig:285.5-8}
\end{figure}

Figures \ref{fig:285.1-3}, \ref{fig:285.4}, and \ref{fig:285.5-8} detail the case where the pump pulses were 1000 ps
long and had peak powers of 400 mW. It was found by trial and error that translation occurred efficiently at a fiber
length of 20 m for pumps with peak powers around 400 mW. It was found that the characteristic time of the optimal HG
basis was $\approx 243$ ps. Figure \ref{fig:285.1-3}(a) shows the absolute value of the $V$ matrix, as defined in
\eqref{eq:GgbSVD}, from  an SVD of the $G_{gb}$ Green function. This Green function corresponds to the case of
translation from the green mode to the blue mode. As discussed previously, the columns of the $V$ matrix represent
the input Schmidt modes, so in this case the $V$ matrix represents the input signal field. That is,
\begin{equation}\label{}
 V_n (\omega) = \sum_j c_j^n \psi_j(\omega).
\end{equation}
The horizontal axis is the Schmidt mode number $n$, where order was determined by the
magnitude of the corresponding singular value $\rho$, largest to smallest, for the first 25
Schmidt modes. The vertical axis is the HG mode number $j$ of the Schmidt modes using the same HG basis used to find the 
Green function. The plotted quantity is $|c_j ^n|$. For example, the first Schmidt mode of the $V$ matrix is
mostly composed of the first HG function and almost completely described by the first four HG functions. Thus we can
conclude that the temporal duration of the first Schmidt mode is close to the characteristic time of the optimal HG
basis, 243 ps.

The output Schmidt modes, which physically are the results of propagating each corresponding input Schmidt mode
through the QFT process in the fiber, are described by the $w$ matrix. The absolute value of this matrix, the column
vectors of which correspond to output in the blue mode, is shown in Fig. \ref{fig:285.1-3}(b). The axes are the same
as those in part (a). Interestingly, the $V$ and $w$ matrices are qualitatively similar, although the $w$ matrix Schmidt 
modes tend to be dominated by either even or odd HG functions whereas the $V$ matrix Schmidt modes are composed of a 
more or less equal number of even and odd HG functions.

The black circles of the red line in Fig. \ref{fig:285.1-3}(c) are the squared singular values $\rho^2$ of the
$G_{gb}$ Green function for this case. The circles of the blue line correspond to the squared singular values
$\tau^2$ of the $G_{gg}$ Green function. The $\rho^2$ describe the translation efficiencies of the input green Schmidt
modes to the output blue Schmidt modes, whereas the $\tau^2$ describe the ``non-translation" efficiencies of the input
green Schmidt modes to the output green Schmidt modes . This corresponds to the transmission and possible reshaping of
the green mode. For the Manley-Rowe relations to be satisfied, the square of these singular values must add to one, 
which they do numerically in this case (as well as in all other cases) to one part in one thousand.\footnote{The 
higher-order singular values are accurate to one part in one thousand, but the lower-order singular values are usually 
more accurate, on the order of one part in one million.} As is evident from the figure, many of the modes translate 
efficiently with the first several having efficiencies of over 90 percent. We call this configuration 
``non-discriminatory" in the sense that it effectively translates many different dissimilar modes efficiently.

Figure \ref{fig:285.4} shows the translation efficiency as a function of length long the fiber for the first three green 
input Schmidt modes. Here translation efficiency is defined as the number of photons in the blue mode divided by the
initial number of photons in the green mode. Note that there were initially no photons in the blue mode. The red, green, 
and blue lines correspond to the first, second, and third Schmidt modes, respectively. The efficiency for these modes is 
very sinusoidal, almost as if the signals were monochromatic and described by \eqref{eq:trans1} and \eqref{eq:trans2}.

Figure \ref{fig:285.5-8} shows the absolute amplitudes and phases in both time and frequency of the first three input 
Schmidt modes (relating to the $V$ matrix) of the green (673 nm) mode. The input amplitudes of the pump fields are also 
shown as a black dashed line. As could be guessed from Fig. \ref{fig:285.1-3}(a), the modes look very much like HG 
functions, although there is some asymmetry and delay in both time and frequency. These effects are likely the results 
of pump evolution, which includes convection, dispersion and CPM. Figure \ref{fig:285.5-8}(a) shows that first few 
Schmidt modes are much narrower in time than the pump fields. Looking at Fig.
\ref{fig:285.5-8}(b), which shows the phases in time, all the input Schmidt modes, disregarding the $\pi$ phase
jumps, are distinctly parabolic with positive curvature. For the phase convention we use, this shape implies the Schmidt 
modes at the fiber's input are down-chirped (frequency decreases in time), meaning that when they propagate through the 
medium, which for the signals is normally-dispersive, they will temporally compress. That is, they are dispersion 
pre-compensated. CPM also chirps the signals in the same way as linear dispersion. Figure \ref{fig:285.5-8}(c) shows 
that the pump fields are nearly monochromatic as compared to the Schmidt modes, although all modes are well within the 
translation phase-matching bandwidth of the fiber.

\subsection{Frequency translation, short pulse}

\begin{figure}[h]
\centering\includegraphics[width=0.8\textwidth]{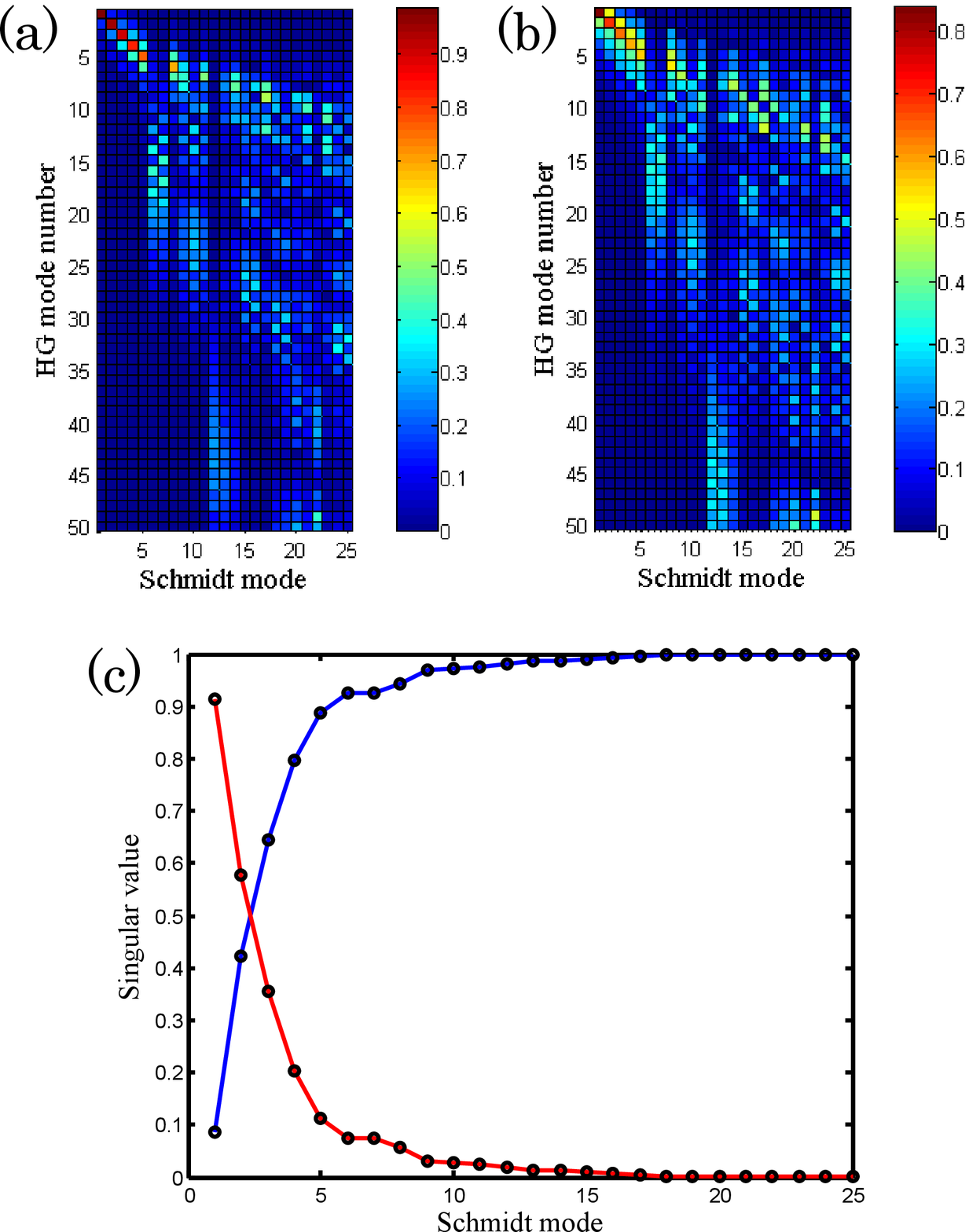}
\caption{Singular values, $V$ matrix, and $w$ matrix of $G_{gb}$ Green function relating to the case with short pumps
(70 ps) having peak powers of 400 mW. (a) Absolute value of the $V$ matrix. (b) Absolute value of the $w$ matrix. (c)
Squared singular values $\rho^2$ (red) and $\tau^2$ (blue) index by the corresponding Schmidt mode number.}
\label{fig:51.1-3}
\end{figure}

\begin{figure}[h]
\centering\includegraphics[width=0.5\textwidth]{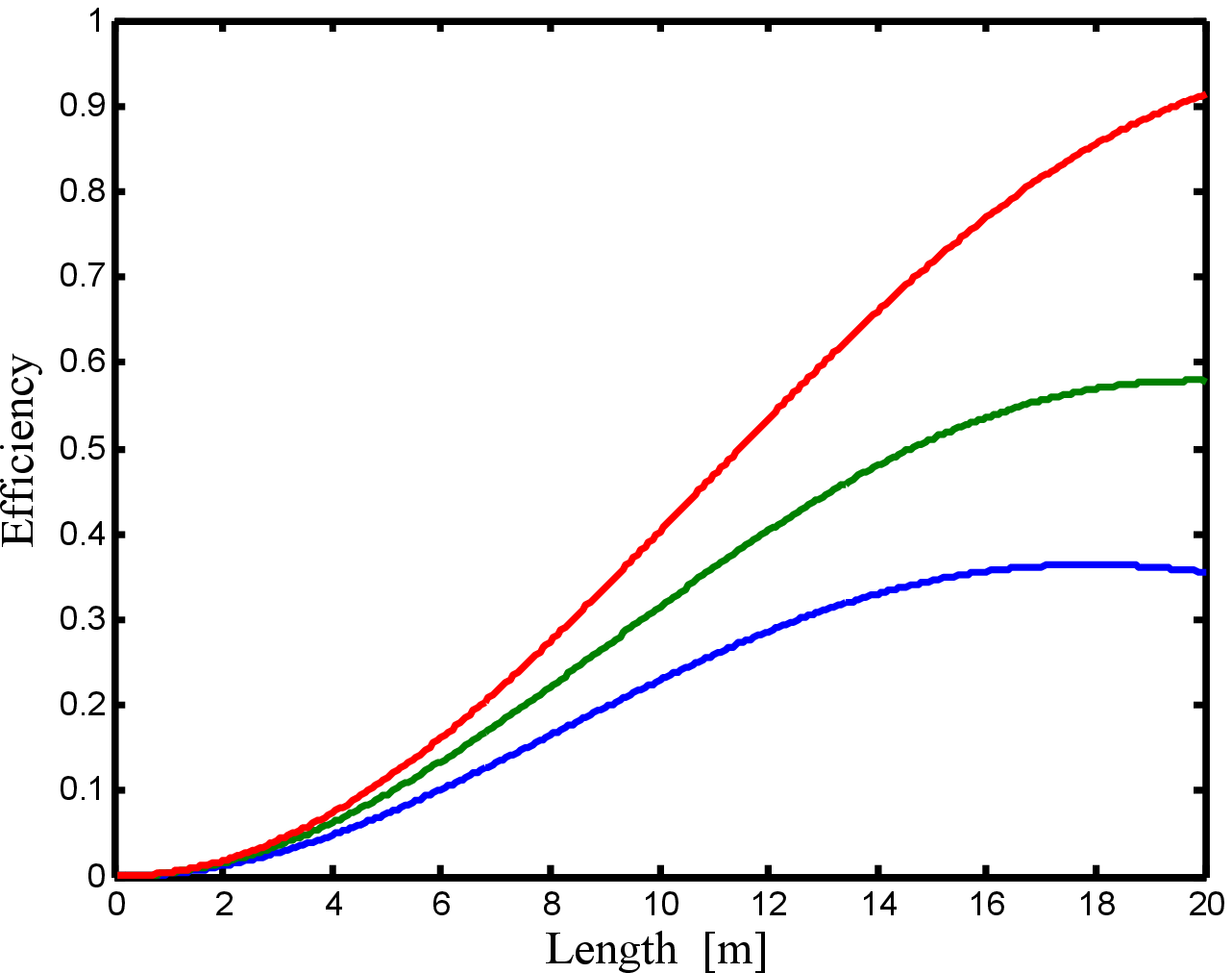}
\caption{Translation efficiency as a function of length along the fiber relating to the case with short pumps (70 ps)
having peak powers of 400 mW. The red, green, and blue lines relate to the first, second, and third green Schmidt 
modes,
respectively.}
\label{fig:51.4}
\end{figure}

\begin{figure}[h]
\centering\includegraphics[width=0.9\textwidth]{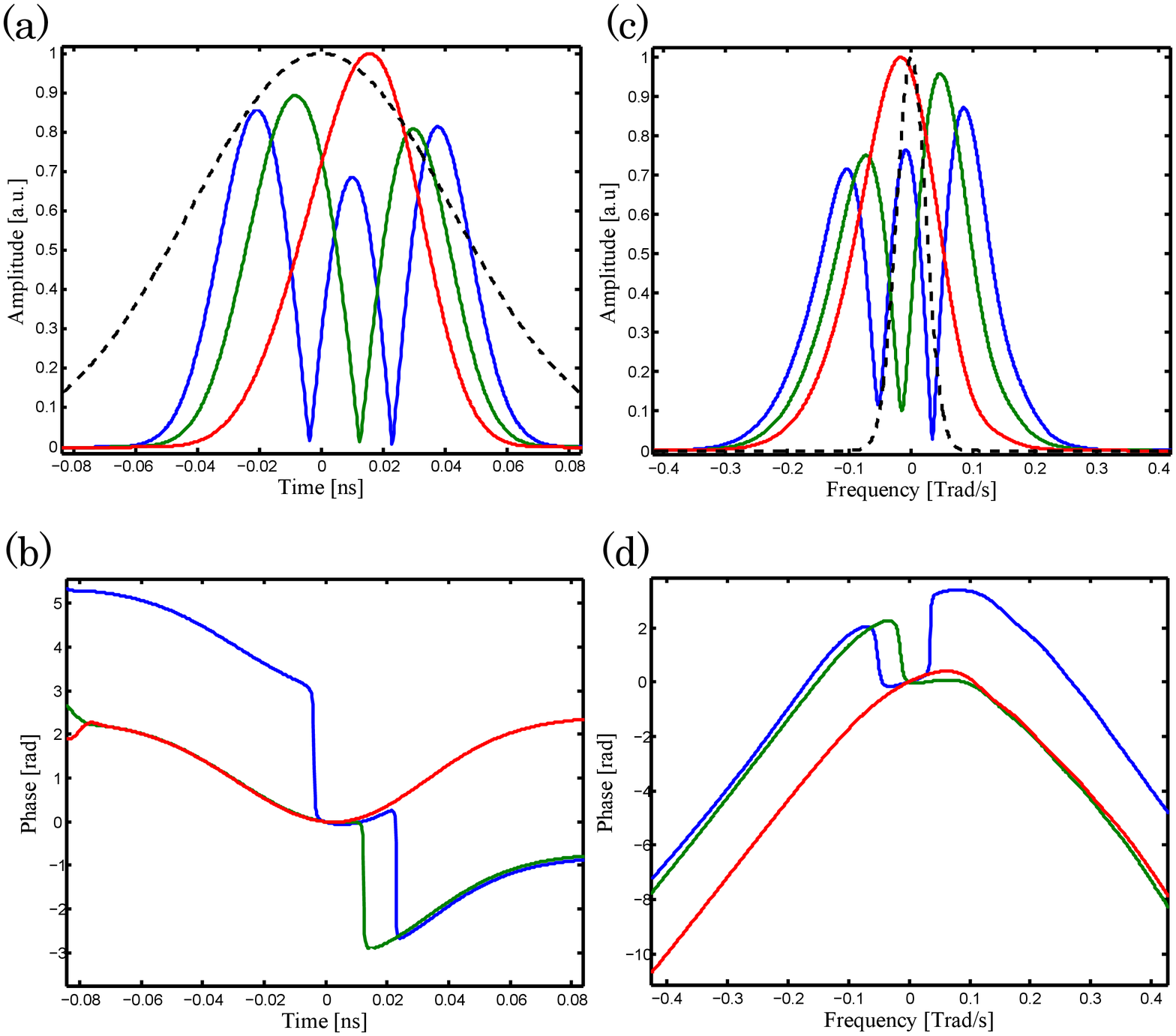}
\caption{Amplitude and phase in both frequency and time of the green 673-nm input Schmidt modes relating to the case 
with short pumps (70 ps) having peak powers of 400 mW.}
\label{fig:51.5-8}
\end{figure}

Figures \ref{fig:51.1-3}, \ref{fig:51.4}, and \ref{fig:51.5-8} detail the case where the pump pulses were 70 ps long
and had peak powers of 400 mW. The goal in studying this case in relation to the former case is to highlight how
using temporally shorter and spectrally broader pump pulses impacts the translation properties of the process. It was
found that the characteristic time of the optimal HG basis for this case was $\approx 43$ ps. Figure
\ref{fig:51.1-3}(a) shows the absolute value of the $V$ matrix (input green Schmidt modes) from the SVD of the $G_{gb}$ 
Green function, while Fig. \ref{fig:51.1-3}(b) shows the absolute value of the $w$ matrix (output Schmidt modes). The 
horizontal and vertical axes have the same meaning as they did in the previous case.

While the absolute values of the $V$ and $w$ matrices are qualitatively similar, they are significantly different
from their counterparts of the previous case. While they start out for low mode order approximately diagonal, like in
the previous case, the Schmidt modes quickly become seemingly random and unordered. This is likely due to these
Schmidt modes' singular values [the squared value of the singular values are shown on the red line of Fig.
\ref{fig:51.1-3}(c)], which quickly drop to almost zero, signifying that almost no translation occurred. Since there
is an infinite number of Schmidt modes with singular values near zero (practically any random combination of HG
modes generally leads to a low singular value), even Schmidt modes of relatively low Schmidt mode number that have
singular values close to zero can look random (non-uniquely defined) in the HG basis. Since in this case only the
first few modes translate with significant efficiency, we call this configuration ``discriminatory" in the sense that
only a few closely related modes effectively translate.

Figure \ref{fig:51.4} shows the translation efficiency as a function of length along the fiber for the first three
input green Schmidt modes in the short-pumps case. The red, green, and blue lines correspond to the first, second, and
third Schmidt modes, respectively. Unlike the former long-pumps case, the efficiency curves of the modes are
drastically different, relating to the fact that their singular values are quite different.

Figure \ref{fig:51.5-8}
shows the absolute amplitudes and phase in both time and frequency of the first three green 673 nm input Schmidt modes 
(relating to the $V$ matrix). The input amplitudes of the pump fields are also shown as a black dashed
line. Qualitatively, the
amplitudes for this case are similar to the amplitudes in the former case; although asymmetric, they are roughly
HG-like. Quantitatively, these amplitudes are much broader spectrally than in the former case. The phases
in time are essentially parabolas with positive curvature much larger than that of the previous case, as expected
because the bandwidths are much larger.

\begin{figure}[b!]
\centering\includegraphics[width=0.9\textwidth]{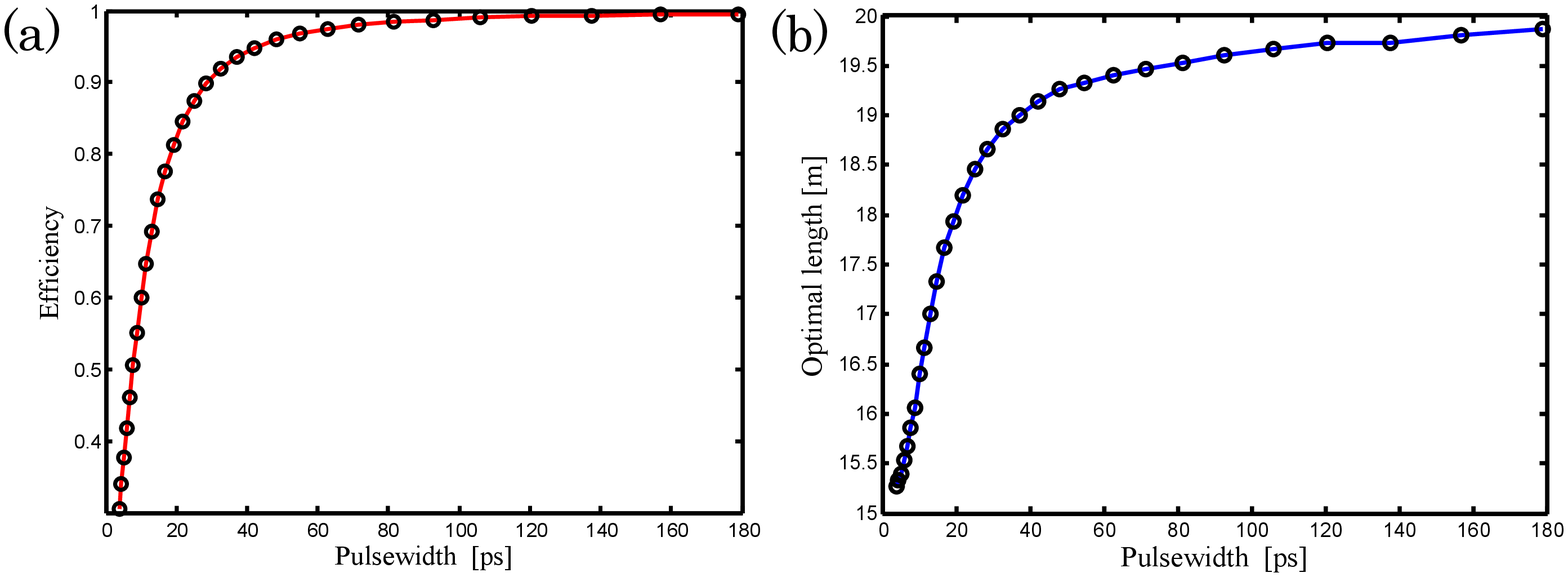}
\caption{Parameter study relating to frequency translation in the case with long pumps (1000 ps) having peak powers
of 400 mW and fiber length 20 m. (a) The maximal translation efficiency achieved in the fiber as a function of signal
pulse width. (b) The length at which the maximal translation efficiency is achieved as a function of signal pulse
width.}
\label{fig:eff.param}
\end{figure}
Figure \ref{fig:eff.param} shows a parameter study where the pump pulses are identical Gaussians with temporal FWHM
durations of 1000 ps and peak powers of 400 mW. In Fig. \ref{fig:eff.param} (a), the maximal translation efficiency
achieved along the fiber is plotted as a function of input signal pulse FWHM duration. For relatively ``long" pulses
in low hundreds of picoseconds, the translation efficiency is quite high due to the near-CW nature of the pump and
the width of the translation bandwidth. As the pulses become shorter, particularly around 43 ps, which corresponds to a 
spectral bandwidth of $92$ Grad/s, the pulse spectral width begins to match that of the translation phase-matching 
bandwidth. Therefore, this dramatic drop in efficiency is likely caused by the signal spectral bandwidth being too large 
for the BS process to accommodate. In Fig. \ref{fig:eff.param}(b), the length at which the maximal translation
occurs is plotted as a function of input signal duration. The curve has the same qualitative shape as the curve in
part (a); for short signal durations, the length at which maximal translation occurs is shorter than that of signals
with long durations. This likely is due to the fact that short pulses have large spectral bandwidths, which because
of CPM from the pumps, become even more spectrally broad as they evolve, making them in harder to
translate. Hence the best translation occurs for these signals when they haven't evolved much. Signals with long
duration and small spectral width do not spread out in frequency as much as the short signals do, so they are less
limited by the magnitude of the translation bandwidth and achieve good translation along the entire fiber.

\begin{figure}[h]
\centering\includegraphics[width=0.9\textwidth]{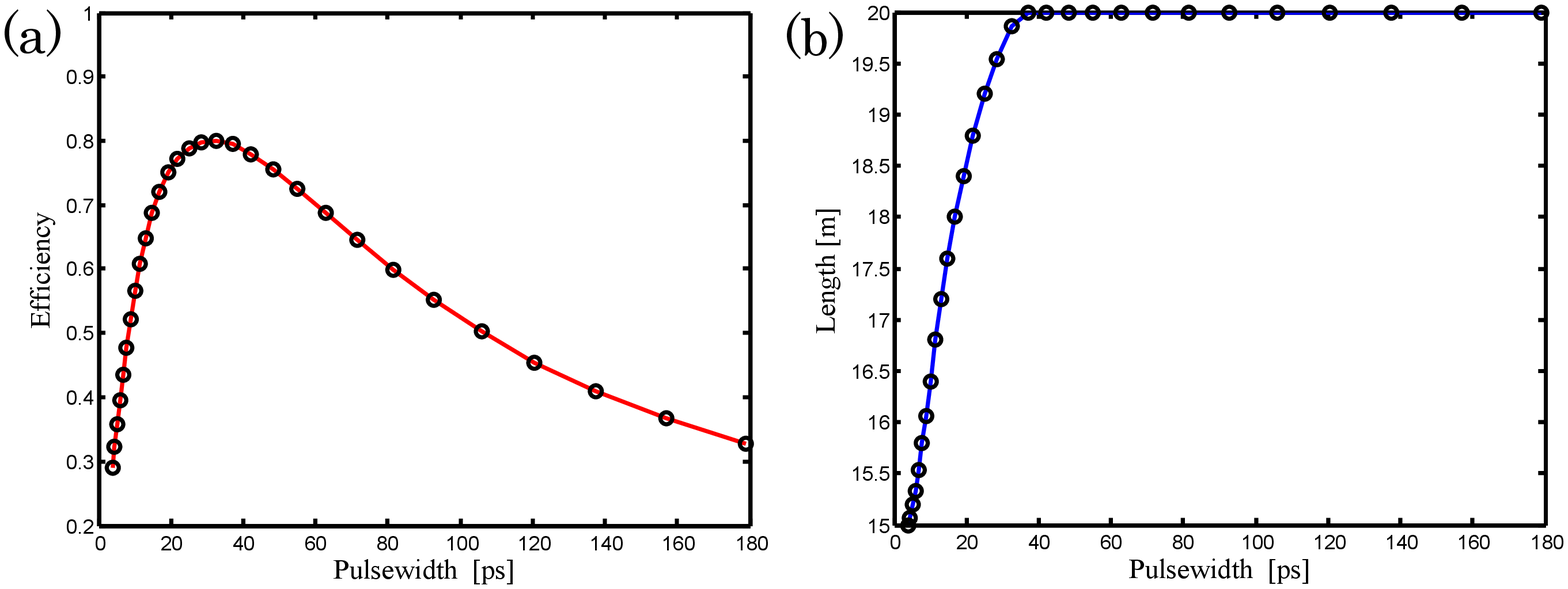}
\caption{Parametric study relating to the case with short pumps (70 ps) having peak powers of 400 mW and fiber length
20 m. (a) The maximal translation efficiency achieved in the fiber as a function of signal pulse width. (b) The
length at which the maximal translation efficiency is achieved as a function of signal pulse width.}
\label{fig:eff.ps70}
\end{figure}

Figure \ref{fig:eff.ps70} show a parameter study where the pump pulses are identical Gaussians with temporal FWHM
durations of 70 ps and peak powers of 400 mW. The efficiency, shown in part (a), for short signal inputs behaves
similarly to the previous long-pumps cases; around 20 ps the efficiency drops sharply due to the signal having larger
translation phase-matching bandwidth than the fiber. Unlike the previous case, the efficiency reaches a maximum
around 35 ps and slowly declines for wider input signal widths. Relatedly, it is at about this signal width that the
length at which the maximal efficiency occurs, shown in part (b), is equal to the length of the fiber, suggesting
that the translation efficiency would be higher for a longer fiber or higher fiber for these signal widths. As the
signal width becomes equal and then greater to the pumps widths there are two main elements affecting the efficiency;
the lack of power/fiber length and the fact that the pump pulses no longer fully encompass the signal pulse in time.
Both of these lead to lower translation efficiency. But it is important to note that for some signal width range,
from about 25 ps to 60 ps, reasonably good translation efficiency will occur for short pump pulses with reasonable
parameters such as peak pump powers of 400 mW and a fiber length of 20 m.

\subsection{HOM interference, long pulse}

\begin{figure}[h]
\centering\includegraphics[width=0.8\textwidth]{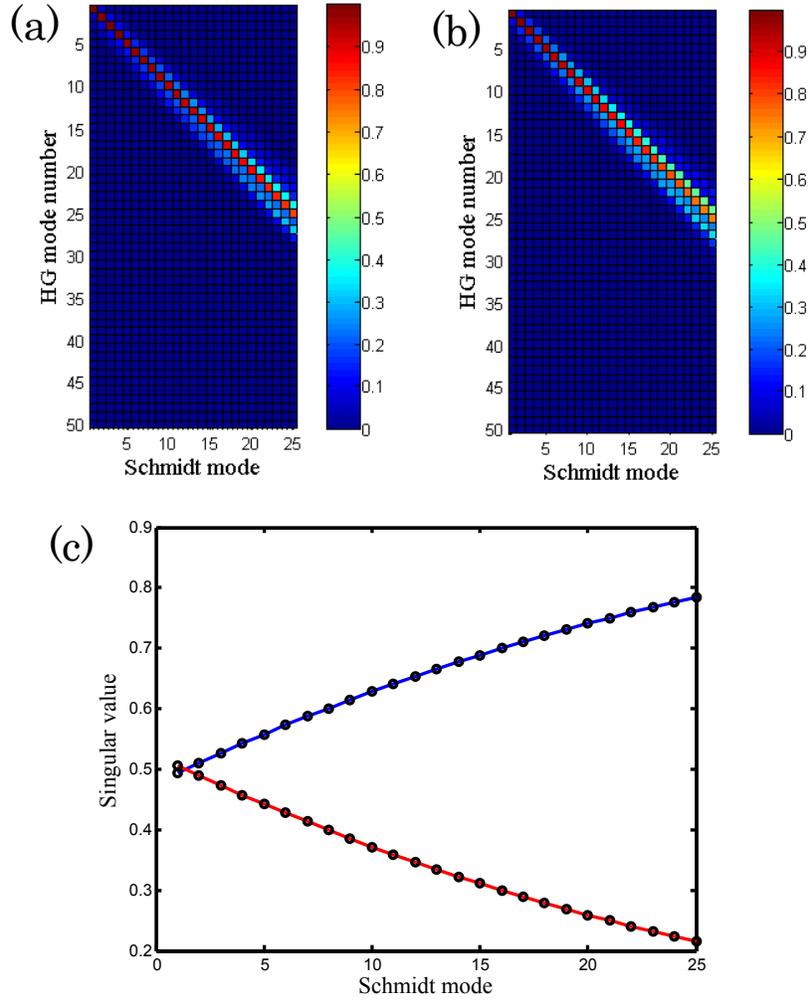}
\caption{Singular values, $V$ matrix, and $w$ matrix of $G_{gb}$ Green function relating to the case with long pumps
(1000 ps) having peak powers of 200 mW optimized for good HOM interference. (a) Absolute value of the $V$ matrix. (b)
Absolute value of the $w$ matrix. (c) Squared singular values $\rho^2$ (red) and $\tau^2$ (blue) index by the
corresponding Schmidt mode number.}
\label{fig:171.1-3}
\end{figure}
\begin{figure}[h]
\centering\includegraphics[width=0.9\textwidth]{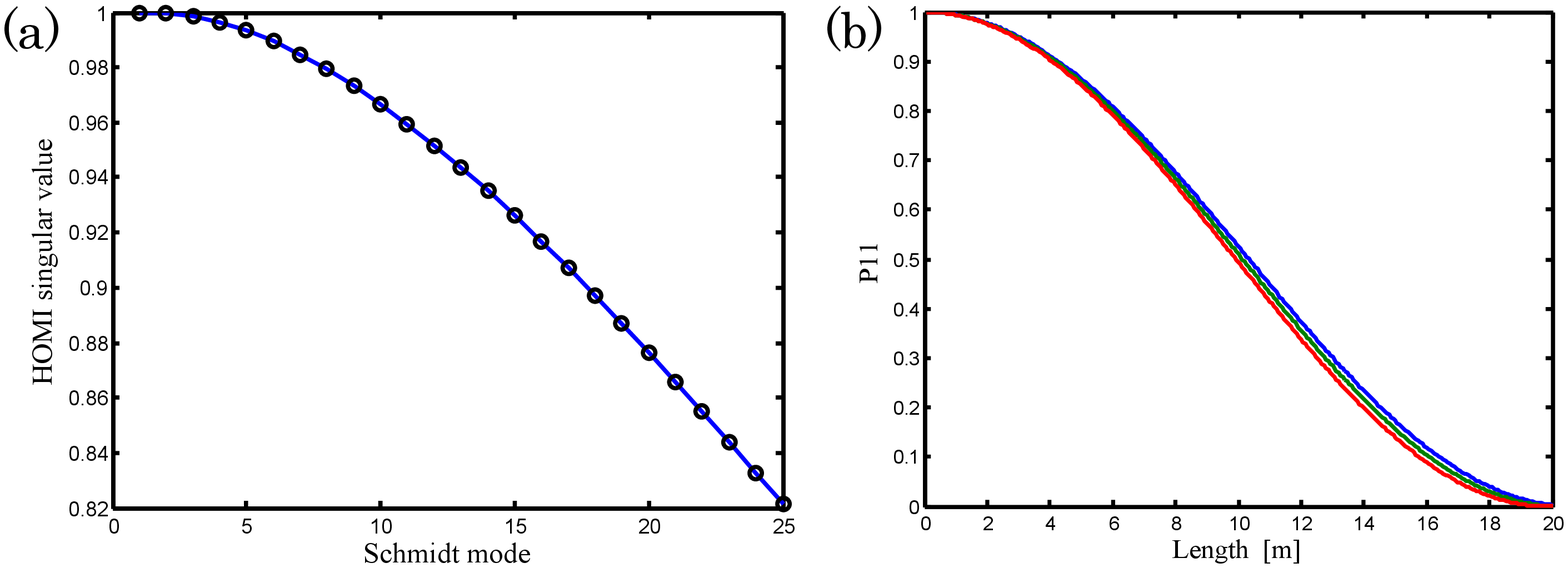}
\caption{HOM singular values $2\rho_n\tau_n$ and $P_{11}$ values relating to the case with long pumps (1000 ps) having
peak powers of 200
mW. (a) HOM singular value index by Schmidt mode number. (b) Two-photon coincidence count probability $P_{11}$ as a
function of length for the first three input Schmidt modes.The red, green, and blue lines relate to
the first, second, and third Schmidt modes, respectively.}
\label{fig:171.4-5}
\end{figure}
\begin{figure}[h!]
\centering\includegraphics[width=0.9\textwidth]{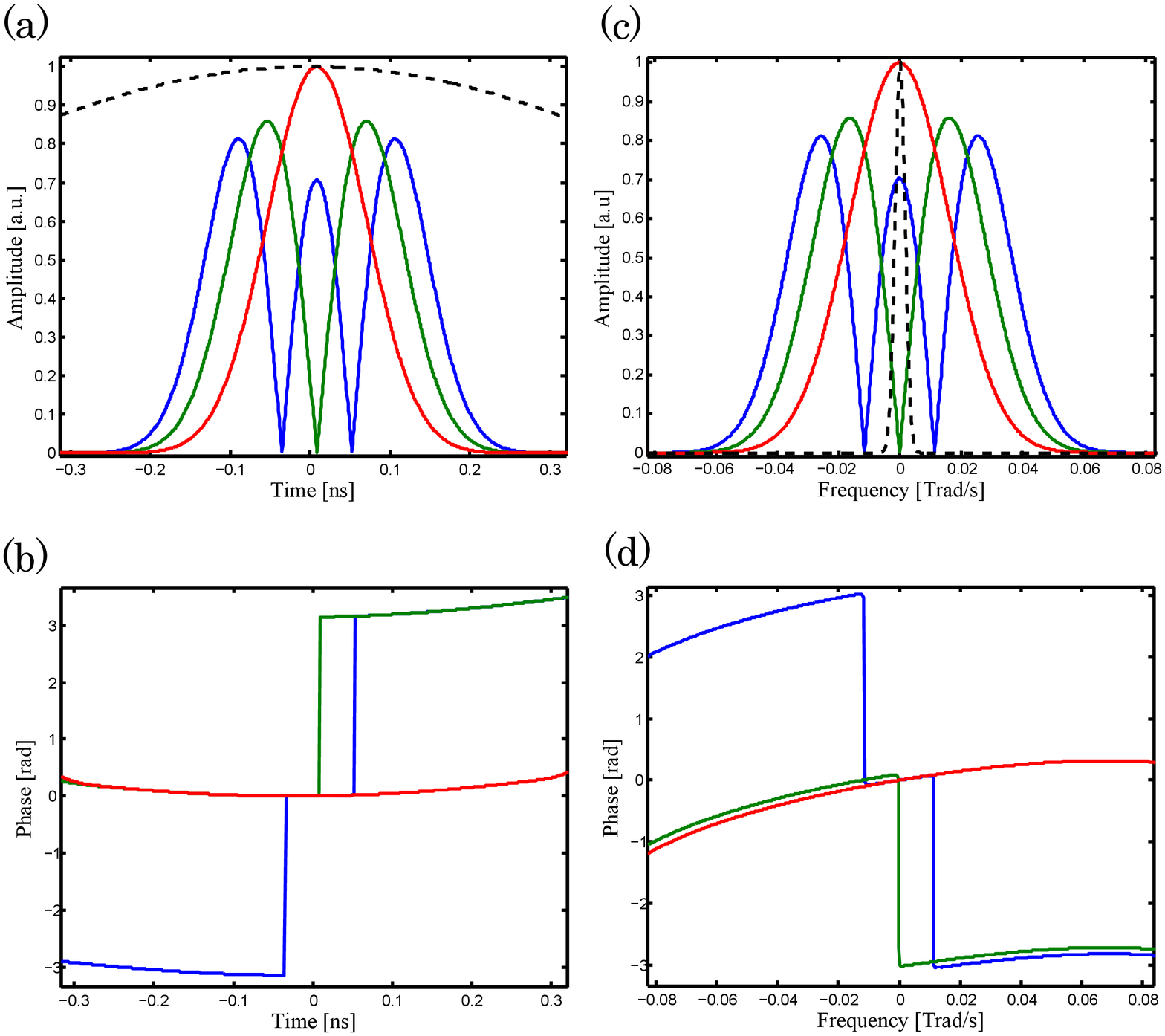}
\caption{Amplitude and phase in both frequency and time of the first three green 673-nm input Schmidt modes relating to 
the case with long pumps (1000 ps) having peak
powers of 200 mW. The red, green, and blue lines relate to the first, second, and third Schmidt modes,
respectively.}
\label{fig:171.6-9}
\end{figure}

Figures \ref{fig:171.1-3}, \ref{fig:171.4-5}, and \ref{fig:171.6-9} detail the case where the pump pulses were 1000
ps long and had peak powers of 200 mW. This configuration was designed to exhibit good two-color HOM interference by
lowering the pumps' peak power so as to lessen the translation efficiency, since the HOM interference effect is
optimal when 50 percent, not 100 percent, translation occurs. It was found that the characteristic time of the
optimal HG basis was $\approx 146$ ps. Figure \ref{fig:171.1-3}(a) shows the absolute value of the $V$ matrix, as
defined in \eqref{eq:GgbSVD}, from  an SVD of the $G_{gb}$ Green function. This Green function corresponds to the
case of translation from the green mode to the blue mode. Figure \ref{fig:171.1-3}(b) shows the absolute value of the
$w$ matrix. The horizontal and vertical axes are the same as they were the previous cases. Surprisingly, both the $V$ 
and $w$ matrices are nearly diagonal, meaning that the input Schmidt mode set and the HG basis functions are nearly 
identical. In Appendix B we present a theoretical model explaining this result. The long pump pulse of 1000 ps enables 
many modes to be translated with nearly similar efficiencies, as can be seen from \ref{fig:171.1-3}(c). For this reason 
we describe this case as ``non-discriminatory" with respect to its two-color HOM interference properties, since many 
different types of modes lead to good HOM interference.

Figure \ref{fig:171.4-5} shows the HOM singular values, as defined by \eqref{eq:HOMIsvd}, as a function of Schmidt
mode number, and the value of $P_{11}$ for the first three input Schmidt modes as a function of fiber length. We
emphasize
that a value $P_{11}$=0 means that two green or two blue photons exit the fiber, but never one green and one blue. This
is
the HOM interference effect as it applies to interference of photons of different colors \cite{Raymer2}. Figure
\ref{fig:171.4-5}(a) shows clearly why this case is non-discriminatory in regards to its HOM interference properties;
many orders of Schmidt modes exhibit HOM singular values near one.

Figure \ref{fig:171.6-9} shows the absolute amplitudes and phase in both time and frequency of the first three green 
input Schmidt modes (relating to the $V$ matrix). The input amplitudes of the pump fields are also shown as a black 
dashed line. Since the $V$ matrix is nearly diagonal, the Schmidt mode amplitudes are almost quantitatively identical to 
the HG basis functions, which stands in contrast to the case in section~\ref{sec:LongPHighP}  which had the same 
duration pulse length but twice the pump power. Similarly to the previous cases, the phase in time is parabolic with 
positive curvature, showing that these inputs are also down-chirped.


\subsection{HOM interference, short pulse}

\begin{figure}[b!]
\centering\includegraphics[width=0.8\textwidth]{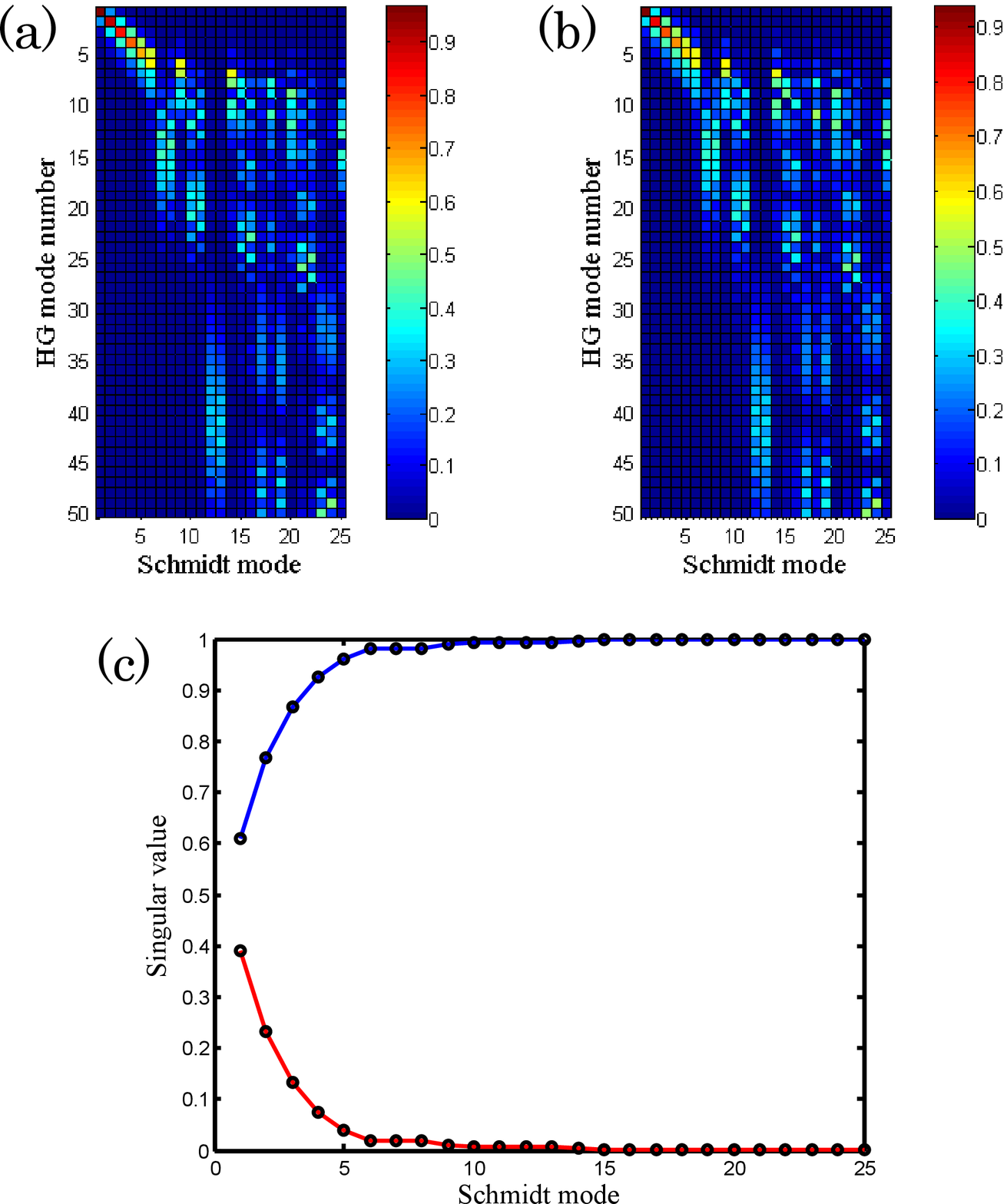}
\caption{Singular values, $V$ matrix, and $w$ matrix of $G_{gb}$ Green function relating to the case with short pumps
(70 ps) having peak powers of 200 mW, optimized for good HOM interference. (a) Absolute value of the $V$ matrix. (b)
Absolute value of the $w$ matrix. (c) Squared singular values $\rho^2$ (red) and $\tau^2$ (blue) indexed by the
corresponding Schmidt mode number.}
\label{fig:45.1-3}
\end{figure}
\begin{figure}
\centering\includegraphics[width=0.9\textwidth]{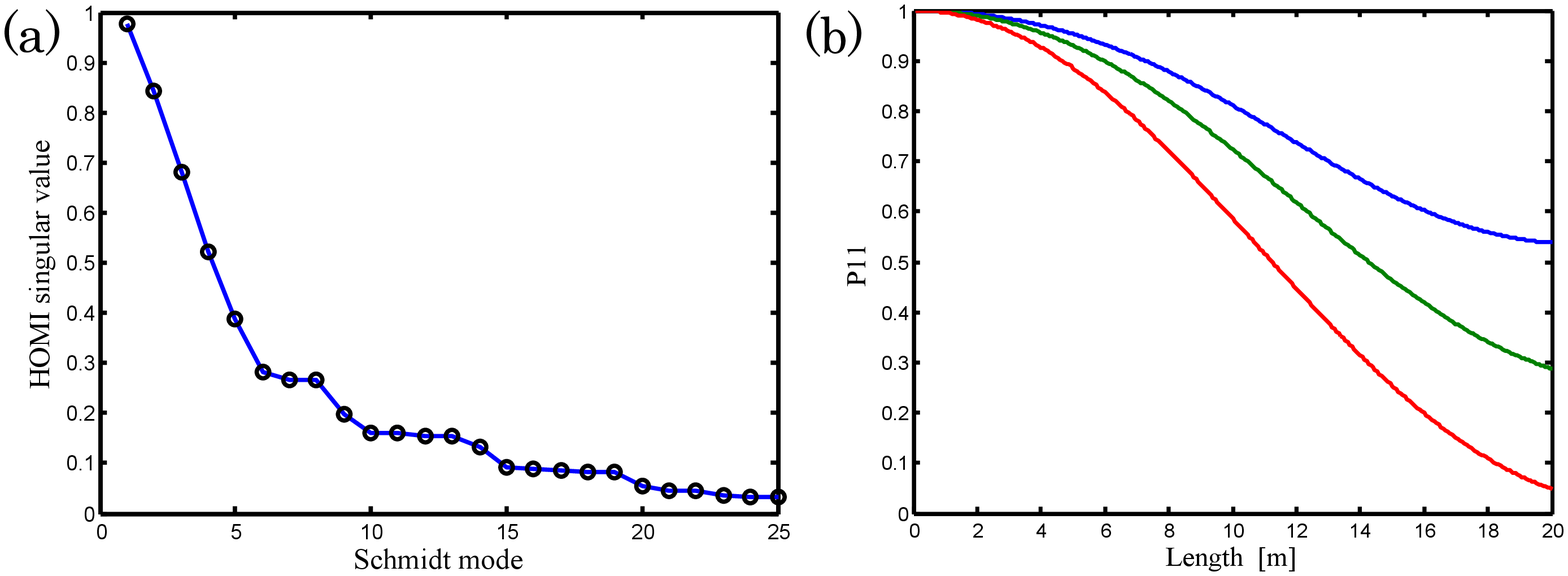}
\caption{HOM singular values and $P_{11}$ relating to the case with short pumps (70 ps) having peak powers of 200 mW. 
(a) HOM singular values indexed by Schmidt mode number. (b) Two-photon coincidence count probability $P_{11}$ values as 
a function of length along the fiber for the first three input Schmidt modes. The red, green, and blue lines relate to 
the first, second, and third Schmidt modes, respectively.}
\label{fig:45.4-5}
\end{figure}
\begin{figure}
\centering\includegraphics[width=0.9\textwidth]{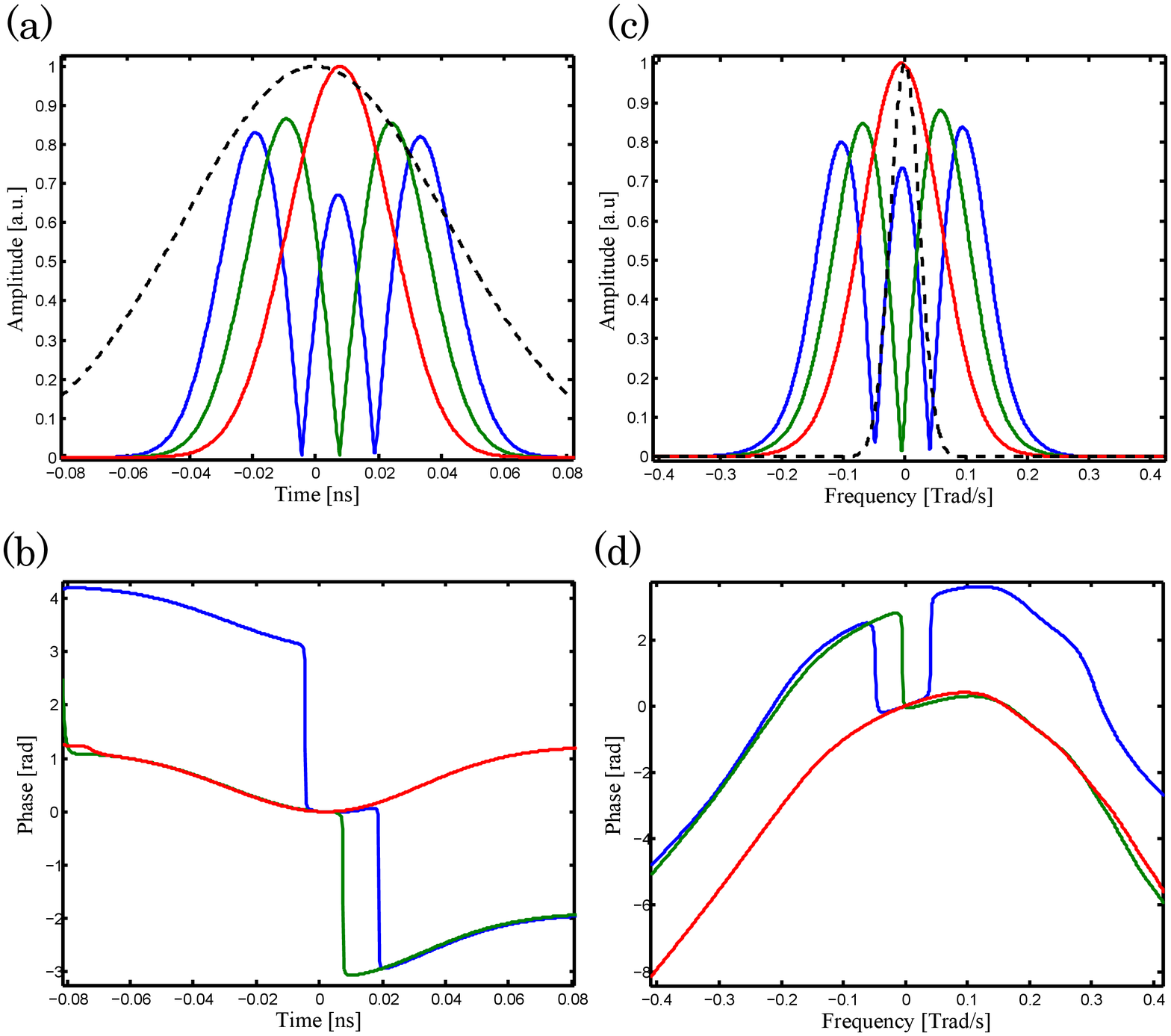}
\caption{Amplitude and phase in both frequency and time of the green 673-nm input Schmidt modes relating to the case 
with short pumps (70 ps) having peak powers of 200 mW. The red, green, and blue lines relate to the first, second, and 
third Schmidt modes, respectively.}
\label{fig:45.6-9}
\end{figure}

Figures \ref{fig:45.1-3}, \ref{fig:45.4-5}, and \ref{fig:45.6-9} detail the case where the pump pulses were 70 ps in
duration and had peak powers of 200 mW, which is close to the optimum for good HOM interference. The goal in studying
this case is to contrast it with the former case that had pumps with the same peak power but were much longer in
duration. This will highlight the relative importance of having spectrally broad and temporally short pump pulses in
regards to the HOM interference behavior. It was found that the optimal characteristic time of the HG basis for this
case was $\approx 38$ ps. Figure \ref{fig:45.1-3}(a) shows the absolute value of the $V$ matrix from the SVD of the
$G_{gb}$ Green function, while Fig. \ref{fig:45.1-3}(b) shows the absolute value of the $w$ matrix. The horizontal
and vertical axes are the same as they were in the previous cases.

\begin{figure}
\centering\includegraphics[width=0.9\textwidth]{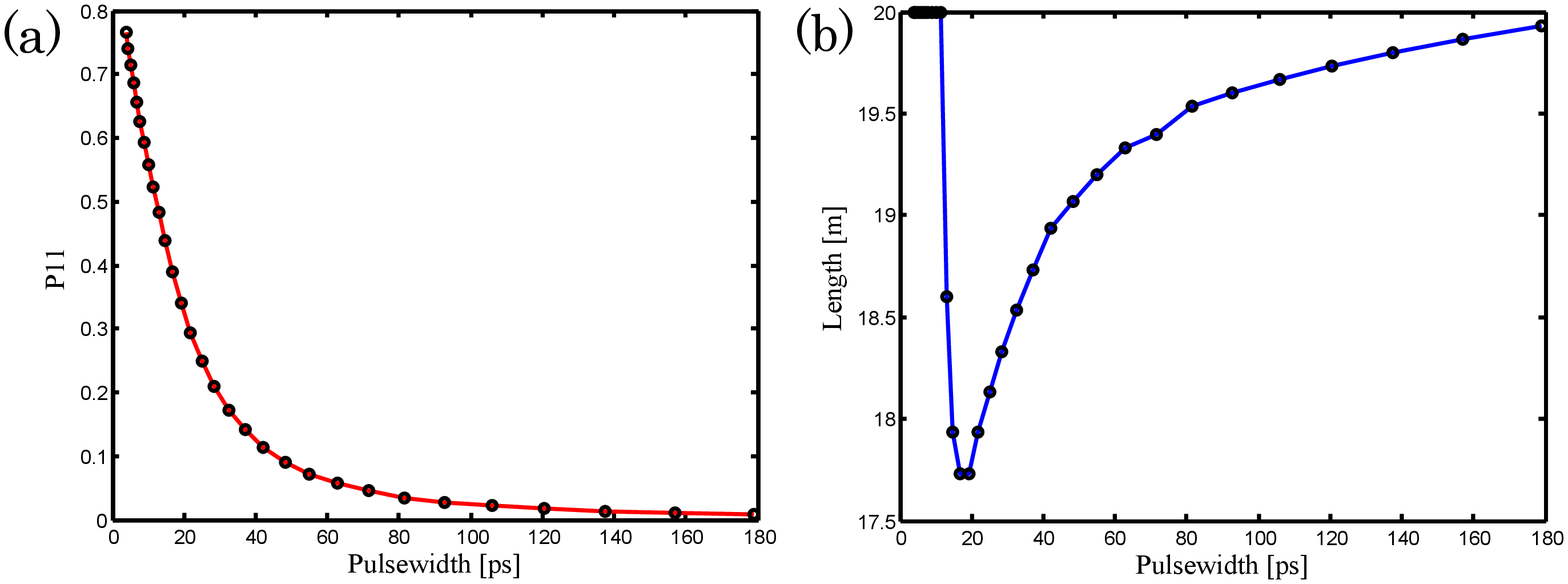}
\caption{Parametric study relating to the case with long pumps (1000 ps) having peak powers of 200 mW. (a) The
minimal $P_{11}$ value achieved in the fiber as a function of signal pulse duration. (b) The length at which the
minimal
$P_{11}$ value is achieved as a function of signal pulse duration.}
\label{fig:P11.param}
\end{figure}


\begin{figure}
\centering\includegraphics[width=0.9\textwidth]{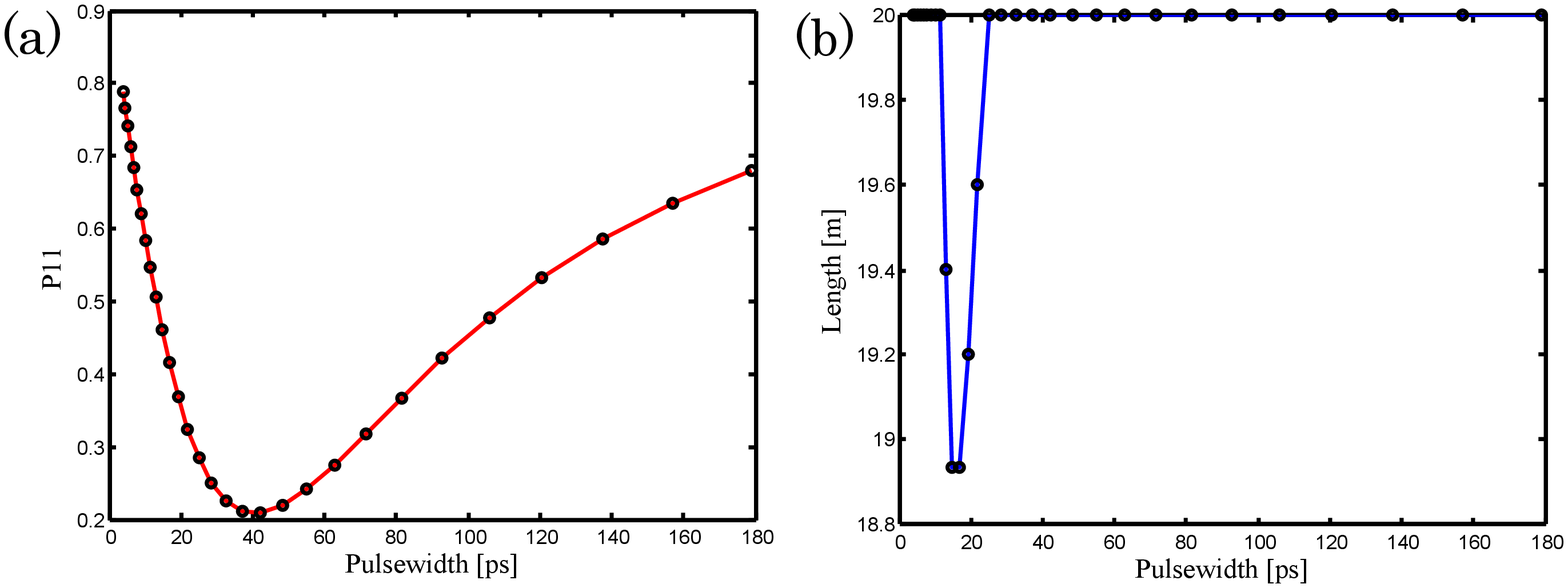}
\caption{Parametric study relating to the case with short pumps (70 ps) having peak powers of 200 mW. (a) The minimal
$P_{11}$ value achieved in the fiber as a function of signal pulse duration. (b) The length at which the minimal 
$P_{11}$ value is achieved as a function of signal pulse duration.}
\label{fig:P11.ps70}
\end{figure}

Like the previous cases, both the $w$ and $V$ matrices are nearly diagonal for low Schmidt mode number, but unlike
the previous case, they quickly become seemingly random combinations of HG functions for higher-order Schmidt modes. The 
reason for this is the same as it was in the other case having short pump pulses; there are many Schmidt modes that have 
low singular values, and practically any combination of higher-order HG functions will have a low, but non-zero singular 
value. As shown in Fig. \ref{fig:45.1-3}(c), only the first few Schmidt modes have a singular value meaningfully above 
zero and will therefore have a more ordered mode structure.

Figure \ref{fig:45.4-5} plots the HOM singular values, as defined by \eqref{eq:HOMIsvd}, as a function of Schmidt
mode and the value of the two-photon coincidence count probability $P_{11}$ for the first three input Schmidt modes as a 
function of fiber length. Figure \ref{fig:45.4-5}(a) shows that, in contrast to the previous case, the HOM singular
values decrease rapidly with Schmidt mode number, making this case highly ``discriminatory" in regards to its HOM
interference properties; only a few Schmidt modes exhibit HOM singular values significantly above zero.

Figure \ref{fig:45.6-9} shows the absolute amplitudes and phase in both time and frequency of the first three green 
input Schmidt modes (relating to the $V$ matrix). The input amplitudes of the pump fields are also
shown as a black dashed line. Similarly to the previous case, these first few Schmidt modes are qualitatively similar to 
the HG basis functions, although they exhibit more asymmetry relative to their width than do the Schmidt modes in the 
previous case. As can be seen in Fig. \ref{fig:45.6-9} (c), the frequency amplitudes have widths that are comparable to 
(and, for the second and third mode, somewhat larger than) the translation bandwidth, which contributes to the fact that 
the higher-order modes experience a low translation efficiency. The phase plots are similar to what they were in the 
previous cases; the phases in time are mostly parabolic with positive curvature, implying the inputs are down-chirped.

Figure \ref{fig:P11.param} shows a parameter study where the pump pulses are identical Gaussians with temporal FWHM 
durations of 1000 ps and peak powers of 200 mW. In Fig. \ref{fig:eff.param} (a), the minimal $P_{11}$ value achieved 
along the fiber is plotted as a function of input signal pulse FWHM duration. For short signal durations the minimum 
$P_{11}$ value hardly changes from 1, which is the initial value ($z=0$) of $P_{11}$ for all modes, but for long signal 
durations the minimum $P_{11}$ value drops to nearly zero. This is most likely due to the short signals having large 
spectral bandwidths and the long signals having small spectral bandwidths as compared to the translation bandwidth. This 
follows because pulses with large relative bandwidth experience poor translation as compared to pulses with small 
relative bandwidths. In Fig. \ref{fig:eff.param}(b), the length at which the minimal $P_{11}$ value occurs is plotted as 
a function of input signal duration. For short signal widths, the optimal length is the same as the fiber length. This 
is likely due to the difficulty translating signals with large bandwidths; it takes the entire length of the fiber to 
translate the small amount of the signal that is within the translation phase-matching bandwidth. This dynamic changes 
once the signal width is on the order of the phase-matching bandwidth, at which point the BS process is no longer 
limited by bandwidth but by the shape and power of the pumps, hence the longer optimal lengths for longer signal 
pulses.

Figure \ref{fig:P11.ps70} shows a parameter study where the pump pulses are identical Gaussians with temporal FWHM
durations of 70 ps and peak powers of 200 mW. The value of $P_{11}$, shown in part (a), for short signal inputs behaves
similarly to the previous long-pumps case: a sharp increase in $P_{11}$ for very short signal widths due to the signal
having larger translation phase-matching bandwidth than the fiber. Unlike the previous case, $P_{11}$ reaches a minimum
around 40 ps and begins to slowly increase for wider input signal widths. The increase for wider signal widths is
likely due to signals beginning or evolving to be wider than the pumps, so that only part of the signals wavepacket
could be translated. The length at which the minimum $P_{11}$ occurs is shown in part (b). This is a qualitatively
similar
result to the previous long-pumps cases. For short signals the optimal length is equal to the length of fiber (fixed
at 20 m), due to the excessive bandwidth of the signal in relation to the phase-matching bandwidth. But when the
signal durations are of the same order as the pumps, the optimal length is not as constrained by the phase-matching
bandwidth in the same manner, and becomes somewhat shorter than the length of the fiber. But when the signals become
as long as or longer than the pump, the overlap between all the fields in time significantly decreases, leading to
less translation and raising the value of $P_{11}$. In this regime the fiber length also limits the translation, hence
the
reason the optimal length is equal to the length of the fiber. This suggests that were the fiber longer the value of
$P_{11}$ would be lower, which indeed is the case.

\section{Analytic derivation of Schmidt modes}

In order to gain further insight into the numerical simulations
presented above, we developed an analytical perturbative solution,
valid for not-too-high conversion efficiencies. The main insights
provided by this solution are: (1) The Schmidt modes are found, to good
approximation, to equal HG functions in this regime. (2)
The origin of the input Schmidt mode temporal delays seen in Figs. \ref{fig:171.6-9} and \ref{fig:45.6-9}
can be understood as arising from differences of group velocities and the
requirement for maximal interaction with the pump pulses in the
fiber.

The detailed derivation is given in Appendix B, the results
of which are summarized here.
Equations (\ref{eq:GgbSVD})--(\ref{eq:GbbSVD}) show how, in general, to express the backward
Green functions as sums of products of the input Schmidt modes $V_n$
and $W_n$, the output modes $v_n$ and $w_n$, and the Schmidt
coefficients $\rho_n$ and $\tau_n = (1 - \rho_n^2)^{1/2}$.
Furthermore, the diagonal Green functions $G_{gg}$ and $G_{bb}$ can
be deduced from the off-diagonal Green functions $G_{gb}$ and
$G_{bg}$ (and vice versa).
In Appendix B we rewrite the third and fourth lines of
(\ref{eq:4mode}) in a frame moving with the average group speed of the
signals. We include the effects of signal convection and second-order
dispersion, but do not include the effects of time-dependent
cross-phase modulation, which were discussed qualitatively after Fig. \ref{fig:285.5-8}. In this frame the dispersion 
relations for
the green and blue signal fields  $k_g(\omega_g) = \beta_1\omega_g
+ \beta_2\omega_g^2/2$ and $k_b(\omega_b) = -\beta_1\omega_b +
\beta_2\omega_b^2/2$ give the associated wavenumbers, correct to
second order in frequency difference. In our simulations $\beta_1$
is negative, so the green mode is faster than the blue mode. For the
parameters in our simulations, the temporal walk-off of pulses is
significant, but pulse broadening by linear group-velocity dispersion
is negligible. Therefore, we set  $\beta_2 = 0$ here for the simplest
analytical solution. We assume the pumps have identical temporal
intensity profiles, proportional to
$\exp ( - t^2 /2\sigma ^2 ) $, where    $\sigma $ is a width parameter. We then find that the
perturbative solution for the signals (to first order in the coupling
parameter $\gamma $) involves Green functions that can be expressed
in SVD form, as are those in (\ref{eq:GgbSVD})--(\ref{eq:GbbSVD}). In this perturbation theory the Schmidt modes are 
independent of the coupling parameter
$\gamma $ and pump peak intensities, but of course this will break
down at higher values of $\gamma $ or pump peak intensities.
Notably, we find that the Schmidt modes are given by HG
functions in this regime. The explicit forms of the input green and
blue Schmidt modes are, respectively:
\begin{eqnarray}
V(\omega _g ) &= &t_0^{1/2} \psi _n (t_0 \omega _g )\exp (+i\omega _g
\beta _1 L/2), \\
W(\omega _b ) &= &t_0^{1/2} \psi _n (t_0 \omega _b )\exp (-i\omega
_b \beta _1 L/2),
\end{eqnarray}
where the $\psi _n (t_0 \omega _g )$ are the HG functions defined in (26), and the temporal scale factor is  $t_0  = 
\left( {0.621\sigma \beta _1 L} \right)^{1/2}$.  Likewise, the output green and blue Schmidt modes are,
respectively:
\begin{eqnarray}
v(\omega _g ) &= &t_0^{1/2} \psi _n (t_0 \omega _g )\exp (-i\omega _g
\beta _1 L/2), \\
w(\omega _b ) &= &t_0^{1/2} \psi _n (t_0 \omega _b )\exp (+i\omega
_b \beta _1 L/2).
\end{eqnarray}
Note that the spectral phases in these mode solutions correspond to temporal
delays or advances, which in this simple model result in both green and blue modes
maximally overlapping with the pump pulses at the midpoint ($z =
L/2$) of the medium. In our simulations $\beta_1$ is negative, so the green input mode is delayed and the blue input 
mode is advanced (although note that in our simulations the two pumps walk off from each other, which is not the case 
for the analytic solution given here).

\begin{figure}[h]
\centerline{\includegraphics[width=2.4in]{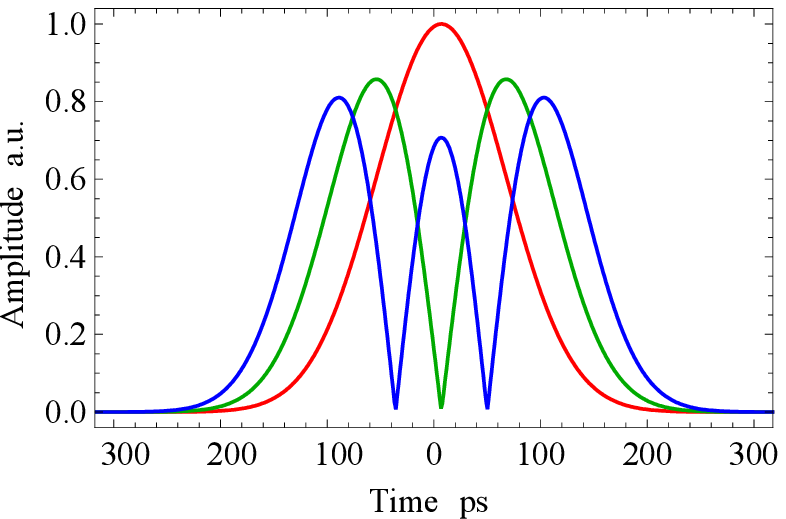}
\hspace{0.1in}
\includegraphics[width=2.4in]{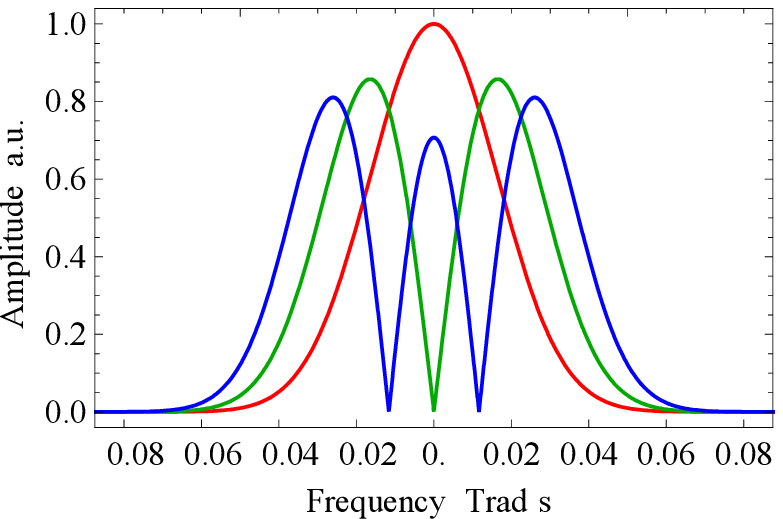}}
\caption{Approximate Schmidt modes for the faster signal plotted as functions of time and
frequency, for long pump pulses. The red, green, and blue lines denote
the first, second and third input Schmidt modes, respectively.} \label{fa1}
\end{figure}

\begin{figure}[h]
\centerline{\includegraphics[width=2.4in]{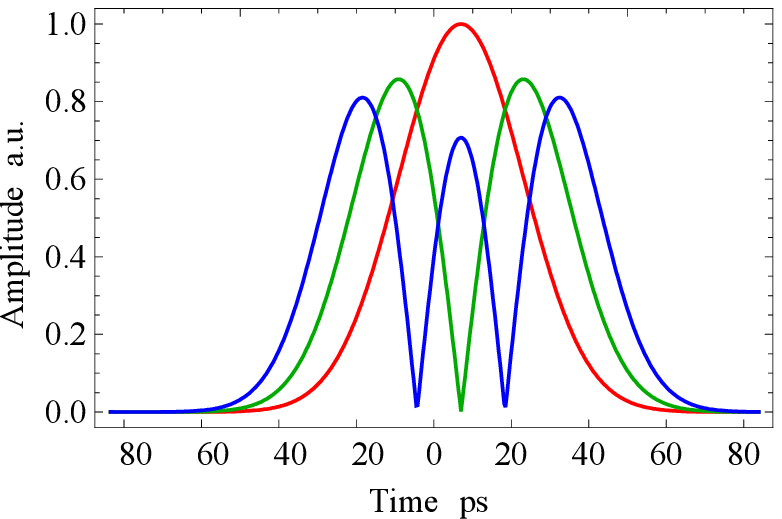}
\hspace{0.1in}
\includegraphics[width=2.4in]{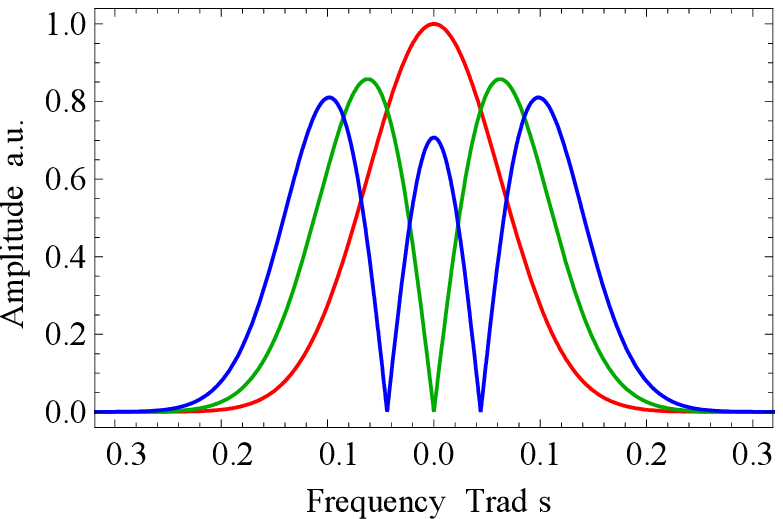}}
\caption{Approximate Schmidt modes for the faster signal plotted as functions of time and
frequency, for short pump pulses. The red, green, and blue lines
denote the first, second and third input Schmidt modes, respectively.}
\label{fa2}
\end{figure}

Using the same parameters as in the numerical simulations leading to Figs. \ref{fig:171.6-9} and \ref{fig:45.6-9}, the 
approximate input Schmidt modes for the blue signal are plotted as functions of time and frequency in Figs. \ref{fa1} 
and \ref{fa2}. (Note that the time-domain modes are also described by HG functions.) For conversion efficiencies up to 
50\%, the approximate analytical solutions predict the modes' widths accurately, and their delays or advances 
qualitatively. For higher conversion efficiencies, accurate analytical solutions are not yet known. Clearly, the 
perturbative solutions will not hold for conversion efficiencies approaching 100\%, as we see significant alterations of 
the modes in Figs. \ref{fig:285.5-8} and \ref{fig:51.5-8}.

\section{Discussion and conclusions}

In this paper we studied quantum frequency translation (QFT) by the four-wave mixing interaction of Bragg scattering 
(BS) using a realistic numerical model of photonic crystal fiber and an analytical approximate solution valid for low 
conversion efficiency. Because BS implements a quantum operator transformation analogous to that of an ordinary beam 
splitter (mode coupler), it is noiseless (background free) \cite{McKinstrie2}. Furthermore, the beam-splitter nature of 
BS leads to linear-optical quantum operations between weak light pulses of different central frequencies, in particular 
two-photon Hong-Ou-Mandel (HOM) interference \cite{Raymer2}. The main questions addressed were: (1) Under what 
conditions does high-efficiency QFT take place? (2) Under what conditions does high-efficiency HOM interference take 
place? (3) Given the shapes of the pump pulses, how can one design ideal signal pulse shapes to optimize the 
above-mentioned processes? (4) How are these ideal signal shapes affected by the level of the conversion efficiency for 
the QFT? That is, are the ideal shapes invariants for a given pump shape, or do they evolve differently at higher 
conversion efficiency than at low conversion efficiency?

Our modeling included realistic input pump and signal fields and effects such as convection, dispersion, self-phase 
modulation and cross-phase modulation. This work extends previous theoretical and numerical work \cite{McKinstrie2, 
Raymer2} on these subjects not only through explicitly modeling of specific cases (specific input signal wavepackets), 
but more generally through the determination of the Green function that governs the BS process for an entire range of 
input signal wave packets. Knowing the Green function for a process enables the calculation of the signal input and 
output Schmidt modes and corresponding singular values (generalizations of eigenmodes and eigenvalues, respectively), 
found by the singular value decomposition of the Green function. Knowledge of the Schmidt modes and their associated 
singular values enables the calculation of the effect of the BS process on the input signals without the need to further 
solve the pulse propagation equations. Most importantly for single-photon inputs, this includes the frequency 
translation and two-color two-photon interference effects. It also allows for the optimization of these effects.

In particular, conditions for achieving high translation efficiency of input signal wave packets with Gaussian pumps
with both long (1000 ps) and short (70 ps) durations at a modest peak power of 400 mW were discovered by finding the
Green function, and then Schmidt modes, of the process. These two cases represent the quasi-CW regime and pulsed
regime, respectively, in regards to the translation phase-matching bandwidth of the process. The long-pumps case was
found to be ``non-discriminatory," where many Schmidt modes were translated with high efficiency. Therefore, this
configuration of fiber and pumps would be useful for effective translation of many different input signal wave
packets or in the case where an input packet is a superposition of the high-efficiency ones. In contrast, the short-pump 
case was found to be ``discriminatory," where only the first few Schmidt modes
experienced significant translation.

Configurations of the discriminatory type would be most suited to applications where it is desirable for only a few
types of wave packets (ideally only one) to experience significant translation, such as in a temporal-mode-selective
add/drop filter \cite{Salehi,Marhic}. Recently it was proposed that such a filter, or ``pulse gate" can be implemented 
using frequency upconversion by sum-frequency generation in an optical crystal driven by a single strong pump pulse 
\cite{Eckstein}. This process is analogous to the BS process studied here, with a major difference being that in 
upconversion the two signal fields being connected by frequency translation must be separated by an amount equal to the 
pump frequency. This limits the process to large frequency separations. The numerical scheme developed here is well 
suited for modeling such a process.

To gain insight into the FWM Bragg scattering process, we developed an analytical perturbation solution. We find that 
for low efficiency QFT the Schmidt modes are well approximated by a set of orthogonal Hermite-Gaussian functions with 
characteristic temporal widths determined simply by group-velocity differences and the spectral width of the pumps. 
Nevertheless, we emphasize that the HG modes become strongly altered for high conversion, in apparent contradiction to 
what is argued in \cite{Eckstein}.

Also studied were the conditions for achieving a highly visible two-color photon interference. We studied two pump
configurations having long (1000 ps) and short (70 ps) durations and a peak power (200 mW) close to the optimum value
for a 20-m-long fiber. Similar to the high-efficiency translation cases, the long-pump case was ``non-discriminatory" 
and the short-pump cases was ``discriminatory" in regards to high visibility of the two-color photon interference 
effect. We verified numerically that for the long-pumps configuration up to about 50 percent conversion efficiency, both 
the input and output Schmidt mode sets are very well approximated by orthogonal Hermite--Gauss (HG) function sets. For 
the short-pumps configuration, the HG functions are reasonable approximations. The approximate theory on which the HG 
functions are based allows one to predict the Schmidt-mode widths and singular values as functions of the physical 
parameters.

In conclusion, QFT by FWM Bragg scattering appears to be a powerful and robust technique for linearly interacting two
quantum fields of distinct frequencies. Besides offering the ability to change the colors of single photons, it may
offer other useful abilities in quantum information science, such as linear-optical quantum computing ``over the
rainbow" of optical and near-IR frequencies.

\section*{Acknowledgments}
This work was partially supported by NSF Grant ECCS-0802109 and by the Oregon Nanoscience and Microtechnologies 
Institute. We thank S. van Enk for helpful comments.


\appendices

\section*{Appendix A: probabilities $P_{20}$ and $P_{02}$}

The probability for the occurrence of the $|2,0\rangle$ and $|0,2\rangle$ states requires care to derive due to the
possibility that degenerate photons would be created in the green and blue frequency regions. In this case the primed
variables attached to the operators cannot be taken over to their unprimed counterparts (as $\omega_{b}^{'}$ to
$\omega_{b}$ was for example). This ambiguity can be avoided by calculating $P_{20}$ directly as the product of the
appropriate bra and ket vectors (similarly for $P_{02}$). From \eqref{eq:HOMIout2} the term corresponding to the
creation of
two green photons leads to an expression for $P_{20}$ of
\begin{equation}\label{eq:P20start}
\begin{split}
    P_{20} &= \langle \psi_{20}||\psi_{20} \rangle = \langle vac| \left(
    \int \int d\bar{\omega}_{g} d\bar{\omega}_{g}^{'} a_{g}(\bar{\omega}_{g})a_{g}(\bar{\omega}_{g}^{'})
A_{gg}^*(L,\bar{\omega}_{g}) A_{gb}^*(L,\bar{\omega}_{g}^{'})
      \right)\\
    &\times \left(
    \int \int d\omega_{g} d\omega_{g}^{'} a_{g}^{\dagger}(\omega_{g})a_{g}^{\dagger}(\omega_{g}^{'})
A_{gg}(L,\omega_{g}) A_{gb}(L,\omega_{g}^{'})
      \right) |vac \rangle\\
      &= \langle vac|
    \int \int \int \int  d\bar{\omega}_{g} d\bar{\omega}_{g}^{'} d\omega_{g} d\omega_{g}^{'}
A_{gg}^*(L,\bar{\omega}_{g}) A_{gb}^*(L,\bar{\omega}_{g}^{'}) A_{gg}(L,\omega_{g}) A_{gb}(L,\omega_{g}^{'}) \\ &
a_{g}(\bar{\omega}_{g})a_{g}(\bar{\omega}_{g}^{'}) a_{g}^{\dagger}(\omega_{g})a_{g}^{\dagger}(\omega_{g}^{'})
     |vac \rangle,\\
\end{split}
\end{equation}
where the variables of the conjugate wavefunction are denoted with an overhead bar. The key to evaluating this
expression is to utilize the field commutation relation. In the frequency variable the commutation relation is
$[a(\omega), a^{\dagger}(\omega^{'})] = \delta(\omega-\omega^{'})$. Hence the commutation relation to be used on the
innermost operators is
\begin{equation}\label{eq:P20delta}
    a_{g}(\bar{\omega}_{g}^{'}) a_{g}^{\dagger}(\omega_{g}) = \delta(\bar{\omega}_{g}^{'} - \omega_{g}) +
     a_{g}^{\dagger}(\omega_{g}) a_{g}(\bar{\omega}_{g}^{'}).
\end{equation}
This breaks the expression into two parts, and leads to the expression, employing dummy variable notation,
\begin{equation}\label{eq:P20}
    P_{20} = \left|\int d\omega A_{gg}^{*}(L,\omega)A_{gb}(L,\omega) \right|^2 + \int d\omega d\omega^{'}
|A_{gg}(L,\omega)|^2 |A_{gb}(L,\omega^{'})|^2.
\end{equation}
The expression for $P_{02}$ is the same when $A_{gg} \rightarrow A_{bb}$ and $A_{gb} \rightarrow A_{bg}$. The two terms
of \eqref{eq:P20} have distinct physical interpretations. The first term, derived in part from the delta function of
\eqref{eq:P20delta}, is essentially the overlap of the two spectral functions $A_{gg}$ and $A_{gb}$, and accounts for
cases in which the two green photons are created in the same mode. This purely quantum mechanical term arises from
the bosonic nature of photons, the operators of which obey the above-mentioned commutation relation. The second term
accounts for the cases in which the two green fields are purely classical, as if these fields did not obey the
quantum bosonic commutation relations.


\section*{Appendix B: derivation of the approximate Schmidt modes}

According to (\ref{eq:GgbSVD})--(\ref{eq:GbbSVD}), the (backward) constituent Green functions are determined by the 
input Schmidt modes $V_n$ and $W_n$, the output modes $v_n$ and $w_n$, and the Schmidt coefficients $\rho_n$ and $\tau_n 
= (1 - \rho_n^2)^{1/2}$. Furthermore, the diagonal Green functions $G_{gg}$ and $G_{bb}$ can be deduced from the 
off-diagonal Green functions $G_{gb}$ and $G_{bg}$ (and vice versa). In this appendix, the off-diagonal Green functions 
and the associated Schmidt modes will be determined approximately, for low conversion efficiencies.

The third and fourth of (24) govern the signal evolution in the time domain. In the frequency domain, the signal 
equations are
\begin{eqnarray} \partial_zA_g(z,\omega_g) &= &ik_g(\omega_g)A_g(z,\omega_g) + i\int \gb(z,\omega_g - 
\omega_b)A_b(z,\omega_b)d\omega_b, \label{a1} \\
\partial_zA_b(z,\omega_b) &= &ik_b(\omega_b)A_b(z,\omega_b) + i\int \gb^*(z,\omega_b - 
\omega_g)A_g(z,\omega_g)d\omega_g, \label{a2} \end{eqnarray}
where $\omega_j$ is an envelope frequency, $k_g(\omega_g) = \beta_1\omega_g + \beta_2\omega_g^2/2$ and $k_b(\omega_b) = 
-\beta_1\omega_b + \beta_2\omega_b^2/2$ are the associated wavenumbers (correct to second order in frequency), and 
$\gb(z,\omega)$ is the Fourier transform of the coupling term $2 \gamma_K A_p^*(z,t)A_q(z,t)$ divided by $2\pi$. 
Equations (\ref{a1}) and (\ref{a2}) are valid in a frame moving with the average group speed of the signals (so 
$\pm\beta_1$ are the relative group slownesses). They include the effects of convection and second-order dispersion, but 
do not include the effect of time-dependent cross-phase modulation (which chirps the signals).

Let $A_j(z,\omega_j) = B_j(z,\omega_j)\exp[ik_j(\omega_j)z]$. Then the transformed amplitudes $B_j$ obey the transformed 
equations
\begin{eqnarray} \partial_zB_g(z,\omega_g) &= &i\int \gb(z,\omega_g - \omega_b)\exp[ik_b(\omega_b)z - ik_g(\omega_g)z] 
B_b(z,\omega_b)d\omega_b, \label{a3} \\
\partial_zB_b(z,\omega_b) &= &i\int \gb^*(z,\omega_b - \omega_g)\exp[ik_g(\omega_g)z - ik_b(\omega_b)z] 
B_g(z,\omega_g)d\omega_g. \label{a4} \end{eqnarray}
In general, the $B$-equations are no simpler than the $A$-equations, because they depend explicitly on $z$. However, in 
the low-conversion-efficiency regime, one can replace the mode amplitudes on the right sides by the input amplitudes. In 
this regime,
\begin{eqnarray} B_g(L,\omega_g) &\approx &B_g(0,\omega_g) + i\int_{-\infty}^\infty\int_0^L 
\kappa(z,\omega_g,\omega_b)B_b(0,\omega_b)dzd\omega_b, \label{a5} \\
B_b(L,\omega_b) &\approx &B_b(0,\omega_b) + i\int_{-\infty}^\infty\int_0^L 
\kappa(z,\omega_b,\omega_g)B_g(0,\omega_g)dzd\omega_g, \label{a6} \end{eqnarray}
where $\kappa(z,\omega_g,\omega_b) = \gb(z,\omega_g - \omega_b)\exp[ik_b(\omega_b)z - ik_g(\omega_g)z]$ is the kernel in 
(\ref{a3}). Note that $\gamma^*(z,\omega) = [\gamma(z,-\omega)]^*$, so $\kappa(z,\omega_b,\omega_g) = 
\kappa^*(z,\omega_g,\omega_b)$.

If the pumps are Gaussian in time and identical, and their relative convection (walk-off) is neglected, then $A_pA_q = 
p_0\exp(-t^2/\tau^2)$, where $p_0$ is the peak pump power and $\tau$ is the (Gaussian) pump width. In this case,
\begin{equation}
\gb(\omega) = [\gamma p_0\sigma(2/\pi)^{1/2}]\exp(-\sigma^2\omega^2/2), \label{a7}
\end{equation}
where the width parameter $\sigma = \tau/2^{1/2}$. Define the wavenumber-mismatch function $\delta(\omega_g,\omega_b) = 
k_g(\omega_g) - k_b(\omega_b)$. Then
\begin{equation} \delta (\omega_g,\omega_b) = \beta_1(\omega_g + \omega_b) + \beta_2(\omega_g^2 - \omega_b^2)/2. 
\label{a8} \end{equation}
By combining the preceding results, and using the identity $\int_0^L e^{-i\delta z}dz = Le^{-i\delta L/2}\sinc(\delta 
L/2)$, one finds that the integrated kernel
\begin{eqnarray} K(\omega_g,\omega_b) &= &[\gamma p_0L\sigma(2/\pi)^{1/2}]\exp[-i\beta_1L(\omega_g + \omega_b)/2 - 
i\beta_2L(\omega_g^2 - \omega_b^2)/4] \nonumber \\
&&\times\exp[-\sigma(\omega_g - \omega_b)^2/2]\sinc[\beta_1L(\omega_g + \omega_b)/2 + \beta_2L(\omega_g^2 - 
\omega_b^2)/4]. \label{a9} \end{eqnarray}%
The $\omega_j$ terms in the first exponential produce delays (advances) in the time domain, whereas the $\omega_j^2$ 
terms produce chirps in the frequency domain (convolutions in the time domain). By making the approximation $\sinc(x) 
\approx \exp(-cx^2/2)$, where $c = 0.3858$, and omitting dispersion (setting $\beta_2 = 0$), which is negligible for the 
parameters of our simulations, one finds that
\begin{eqnarray} K(\omega_g,\omega_b) &\approx &[\gamma p_0 L\sigma(2/\pi)^{1/2}]\exp[-i\beta_1L(\omega_g + \omega_b)/2] 
\nonumber \\
&&\times\exp[-\sigma^2(\omega_g - \omega_b)^2/2 - \beta^2(\omega_g + \omega_b)^2/2], \label{a10} \end{eqnarray}
where the walk-off parameter $\beta = c^{1/2}\beta_1L/2$.

Because kernel (\ref{a10}) is a Gaussian function of frequency, it can be decomposed using the Mehler identity 
\cite{Mehler,Morse}
\begin{equation}
\exp\biggl[-{(1 + \mu^2)(x^2 + y^2) \over 2(1 - \mu^2)} + {2\mu xy \over (1 - \mu^2)}\biggr] = [\pi(1 - 
\mu^2)]^{1/2}\sum_{n=0}^\infty |\mu|^n\psi_n(x)\psi_n(y), \label{a11}
\end{equation}
where the Hermite-Gaussian functions $\psi_n$ were defined in (27). By comparing (\ref{a10}) and (\ref{a11}), one finds 
that $\mu = (\sigma - \beta)/(\sigma + \beta)$, $x = t_g\omega_g$ and $y = t_b\omega_b$, where $t_g = 
(2\beta\sigma)^{1/2} = t_b$. Hence, kernel (\ref{a10}) has the singular value (Schmidt) decomposition
\begin{equation}
K(\omega_g,\omega_b) = \sum_{n=0}^\infty \lambda_n\phi_n(\omega_g)\phi_n(\omega_b) \exp[-i\beta_1L(\omega_g + 
\omega_b)/2], \label{a12}
\end{equation}
where the Schmidt coefficients $\lambda_n = \gamma p_0L[(\sigma/\beta)(1 - \mu^2)]^{1/2}|\mu|^n$ and the (normalized) 
Schmidt modes $\phi_n(\omega_j) = t_j^{1/2}\psi_n(t_j\omega_j)$. The kernel is separable when $\mu = 0$ ($\sigma = 
\beta$), in which case $\lambda_0 = \gamma p_0L$ and $t_j = 2^{1/2}\sigma = \tau$. In this case, only one mode will be 
partially translated; the others will be unaffected, creating a mode-selective filter, similar to that proposed in 
\cite{Eckstein}.

Let $H_{gb}(\omega_g,\omega_b)$ be the Green function that describes the effect on the output transformed amplitude 
$B_g(L,\omega_g)$ of the input transformed amplitude $B_b(0,\omega_b)$. Then (\ref{a5}) implies that 
$H_{gb}(\omega_g,\omega_b) = iK(\omega_g,\omega_b)$. The associated (forward) Green function for the original 
amplitudes, $A_g(L,\omega_g)$ and $A_b(0,\omega_b)$, is $J_{gb}(\omega_g,\omega_b) = 
H_{gb}(\omega_g,\omega_b)\exp(i\beta_1L\omega_g)$. Hence,
\begin{equation}
J_{gb}(\omega_g,\omega_b) = i\sum_{n=0}^\infty \lambda_n\phi_n(\omega_g)\phi_n(\omega_b)
\exp[i\beta_1L(\omega_g - \omega_b)/2], \label{a13}
\end{equation}
from which it follows that the output (green) and input (blue) Schmidt mode functions are given by
\begin{eqnarray}
v(\omega_g) &= &\phi_n(\omega_g)\exp(-i\beta_1L\omega_g/2), \label{a14} \\
W(\omega_b) &= &\phi_n(\omega_b)\exp(-i\beta_1L\omega_b/2), \label{a15}
\end{eqnarray}
respectively. Likewise, (\ref{a6}) implies that the transformed Green function $H_{bg}(\omega_b,\omega_g) = 
iK^*(\omega_g,\omega_b)$, and the original Green function $J_{bg}(\omega_b,\omega_g) = 
H_{bg}(\omega_b,\omega_g)\exp(-i\beta_1L\omega_b)$. Hence,
\begin{equation}
J_{bg}(\omega_b,\omega_g) = i\sum_{n=0}^\infty \lambda_n\phi_n(\omega_b)\phi_n(\omega_g)
\exp[-i\beta_1L(\omega_b - \omega_g)/2], \label{a16}
\end{equation}
from which it follows that the output (blue) and input (green) Schmidt mode functions are given by
\begin{eqnarray}
w(\omega_b) &= &\phi_n(\omega_b)\exp(+i\beta_1L\omega_b/2), \label{a17} \\
V(\omega_g) &= &\phi_n(\omega_g)\exp(+i\beta_1L\omega_g/2), \label{a18}
\end{eqnarray}
respectively. To obtain these, we needed to account for the distinction between forward and backward Green functions, 
and note that here we are propagating $A$, which corresponds to the quantum operator $a$, not $a^{\dagger}$. If 
$\beta_1$ is positive, the green mode is slower than the blue mode. The preceding results show that the green input is 
advanced and the blue input is delayed, so their collision is centered on the midpoint of the fiber. This location 
maximizes the distance over which the signals interact with the peaks of the pump pulses. In our simulations $\beta_1$ 
is negative, so the green input is delayed and blue input is advanced.

\end{document}